\documentclass{nature}

\usepackage[utf8]{inputenc} 
\usepackage[T1]{fontenc}    
\usepackage{hyperref}       
\usepackage{url}            
\usepackage{booktabs}       
\usepackage{nicefrac}       
\usepackage{microtype}      
\usepackage{lipsum}
\usepackage{amsmath}
\usepackage{amssymb}
\usepackage[normalem]{ulem}
\usepackage{lineno}
\usepackage{longtable}
\usepackage{booktabs}

\usepackage{caption}
\usepackage{svg}

\usepackage{graphicx}
\makeatletter
\let\saved@includegraphics\includegraphics
\AtBeginDocument{\let\includegraphics\saved@includegraphics}
\renewenvironment*{figure}{\@float{figure}}{\end@float}
\makeatother

\title{Stress-testing the Resilience of the Austrian Healthcare System Using Agent-Based Simulation}

\author{Michaela Kaleta$^{1,2,\dagger}$, Jana Lasser$^{1,3,\dagger}$, Elma Dervic $^{1,2}$, Liuhuaying Yang $^{1,2}$, Johannes Sorger $^{1,2}$, Ruggiero Lo Sardo $^{1,2}$, Stefan Thurner $^{1,2}$, Alexandra Kautzky-Willer $^{4,5}$, Peter Klimek$^{1,2,*}$}

\date{\today} 

\begin{document}

\maketitle

\begin{affiliations} 
\item Complexity Science Hub Vienna, Josefst\"adter Stra\ss e 39, 1080 Vienna, Austria;
\item Medical University of Vienna, Section for Science of Complex Systems, CeMSIIS, Spitalgasse 23, 1090 Vienna, Austria;
\item Graz University of Technology; Institute for Interactive Systems and Data Science, Inffeldgasse 16C, 8010 Graz, Austria;
\item Department of Internal Medicine III, Clinical Division of Endocrinology and Metabolism, Medical University of Vienna, W\"ahringer G\"urtel 18–20, A-1090 Vienna, Austria;
\item Gender Institute, A-3571 Gars am Kamp, Austria
\end{affiliations}

$^{\dagger}$Contributed equally

$^*$Correspondence: peter.klimek@meduniwien.ac.at

\begin{abstract}
Patients do not access physicians at random but rather via naturally emerging networks of patient flows between them.
As retirements, mass quarantines and absence due to sickness during pandemics, or other shocks thin out these networks, the system might be pushed closer to a tipping point where it loses its ability to deliver care to the population.
Here we propose a data-driven framework to quantify the regional resilience to such shocks of primary and secondary care in Austria via an agent-based model.
For each region and medical specialty we construct detailed patient-sharing networks from administrative data and stress-test these networks by removing increasing numbers of physicians from the system.
This allows us to measure regional resilience indicators describing how many physicians can be removed from a certain area before individual patients won't be treated anymore.
We find that such tipping points do indeed exist and that regions and medical specialties differ substantially in their resilience.
These systemic differences can be related to indicators for individual physicians by quantifying how much their hypothetical removal would stress the system (risk score) or how much of the stress from the removal of other physicians they would be able to absorb (benefit score).
Our stress-testing framework could enable health authorities to rapidly identify bottlenecks in access to care as well as to inspect these naturally emerging physician networks and how potential absences would impact them.
\end{abstract}

\newpage

\maketitle

\section{Introduction}

Practising physicians form the backbone of our healthcare systems to provide the population with access to health care, i.e. "the timely use of personal health services to achieve the best health outcomes"~\cite{InstMed1993}.
Access to health care might be hampered by structural barriers including the number, type, concentration and location of care providers, but also by how quickly they can be accessed. ~\cite{AHCRQ2018}
Consequently indicators that seek to quantify the access to health care are often derived from these metrics, such as the number of care providers per capita in a given region~\cite{WHO2022}.
It was recently shown that indicators such as the density of care providers fall short of capturing structural barriers to health care that arise from the fact that patients do not access physicians at random~\cite{LoSardo2019}.
Rather, physicians are embedded in informal and emergent networks of flows of patients between them~\cite{Pham09, Lee11, Landon12, Landon13, Sauter14, Duftschmid19}.
These networks might, for instance, emerge because of geographic proximity, of one physician tending to recommend another one, or due to physicians acting as holiday substitutes to each other, and thereby encode how likely a patient will access a given care provider given that s/he has already accessed a specific other one.
Such network-derived indicators for access to health care can paint completely different pictures on systemically important physicians with respect to quantitative indicators that solely focus on the density of care providers~\cite{LoSardo2019}.

The functionality of the healthcare system in many developed countries is increasingly challenged by demographic shifts both in the patient population~\cite{Beard15} and the health workforce~\cite{Silver16} (retirement waves), climate-related risks such as extreme weather~\cite{Costello2013, Salas2019} as well as future and currently emerging infectious diseases~\cite{Rawaf2020}.
Yet, surprisingly little attention is given to the problem of quantifying and stress-testing the resilience of national healthcare systems, as opposed to, e.g., economic or financial systems where such stress tests are a common risk management practice~\cite{Adrian2020}.
In this work we aim to fill this gap by developing a stress-testing framework to quantify the resilience of primary and secondary care by physicians in Austria using agent-based simulations.
We adopt the notion of resilience as the ability to manage and regulate the adaptive capacities of a complex adaptive system that can be described via networks~\cite{Woods15}.
In particular, we consider how regional access to primary and secondary care by physicians depends on the capacities of individual physicians when confronted with shocks that reduce the number of available physicians (be it through retirement, quarantine, or other external shocks). We model the system's response to such a shock in terms of a restructuring of the emergent patient-sharing network and quantify how likely patients are to find new care providers with sufficient capacity within a specific location and time period.
The agent-based model is fully calibrated to observational health care access data that covers approx. 100 million visits of 7,630,498 patients at 9,580 physicians of 13 different specialties in Austria.

Our approach can be briefly outlined as follows. Initially, all physicians are available. We consecutively remove individual physicians of a certain specialty and simulate the ensuing re-distribution of patients within the healthcare system as patients of the removed physicians try to find a new health care provider. Patients choose their new physicians with probabilities that are proportional to the number of shared patients between the removed and the potential new physician. Based on their remaining free capacity, the newly chosen physicians will either accept some additional patients or reject them. The rejected patients will continue to look for a new available physician in the next time step, until they reach the maximum number of physicians they are willing to contact (or a maximum distance they are willing to travel). If patients have not been accepted by a given number of physicians, these patients give up on getting an appointment and become 'lost'.  We assume the patients' starting locations to be the municipality of their initial physician. The one-by-one removal process of physicians continues until all physicians become unavailable. 

In this work we derive two sets of indicators to measure the regional resilience of primary and secondary care sectors. One set of indicators quantifies properties related to the resilience of the entire sector (e.g. specialty) in a given region, for example internal medicine in Vienna. The second set of indicators quantifies properties of individual physicians. For each physician we, therefore, define a risk and a benefit score. The risk score of a physician quantifies how much the removal of that physician affects other physician's capacities (the higher this score, the more likely it is that other physicians will exceed their capacity should the given physician be removed). The benefit score, on the other hand, relates to the free capacity of a physician.

By considering the percentage of physicians with a given specialty in a region that have a risk or benefit score higher than the nation-wide average, we compute regional risk and benefit levels for each specialty. Furthermore, for each region and specialty we measure the free capacity (defined as the percentage of physicians that can be removed before 20\% of the initially free capacity has been filled up) and an indicator for the number of lost patients (percentage of physicians to be removed before 1\% of the patients in a region don't find a new care provider in the model).

\begin{figure*}
    \centering
    \includegraphics[width=1\textwidth]{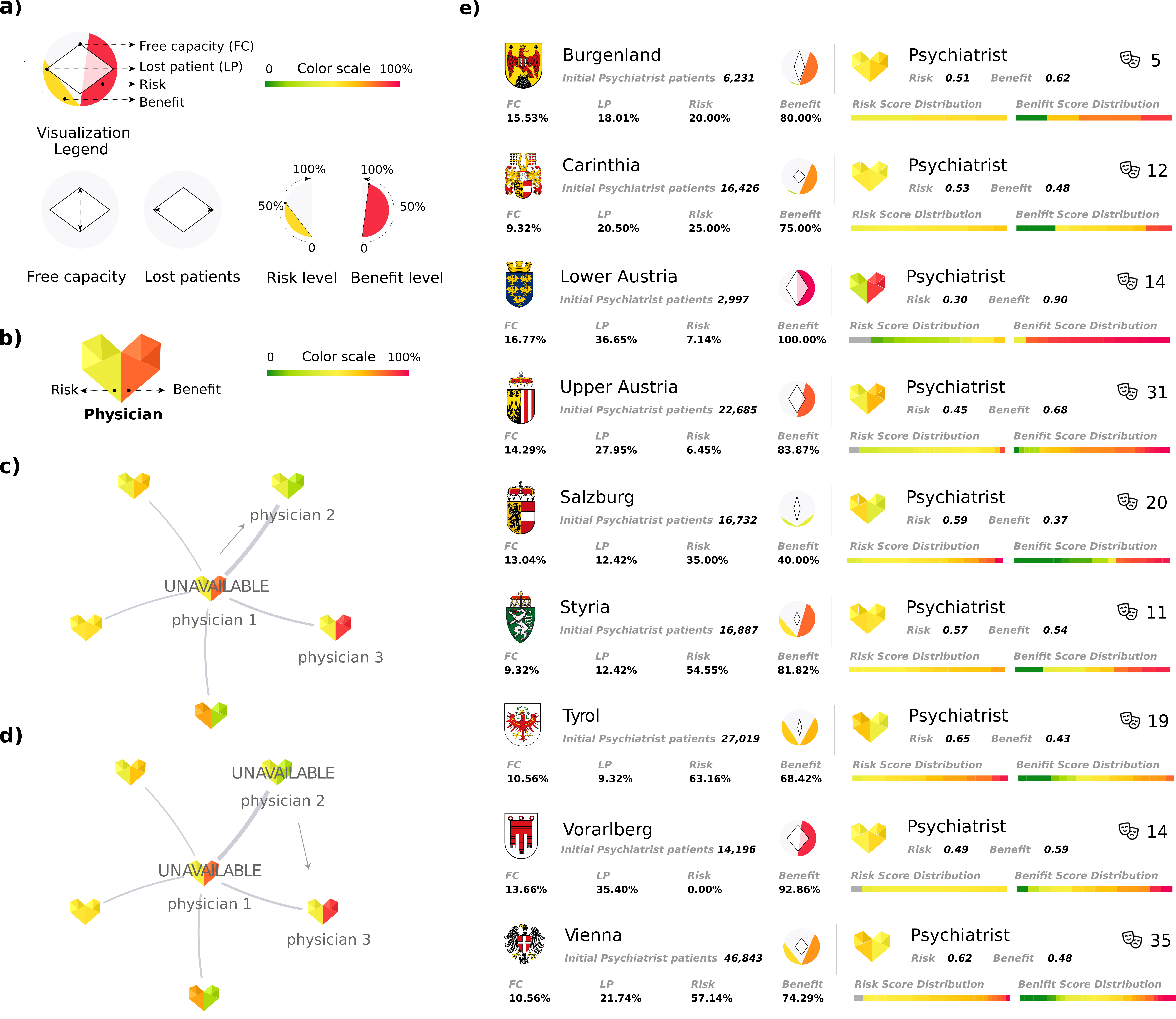}
    \caption{Schematic overview. a) Custom glyph used to describe state-level results. b) Custom glyph for aggregated specialty- or  physician-specific resilience indicators. c) and d) Example of physicians' patient sharing network and patient displacement: (c) physician 1 becomes unavailable and a share of their patients try to re-locate to  physician 2. As a result, physician 2 becomes unavailable as well (d). (e) State-specific result panel for selected medical specialties (psychiatrists) including averaged resilience indicators, numbers of patients and physicians and distributions of risk and benefit scores.} 
    \label{fig:model}
\end{figure*}

Agent-based models often produce complex and high-dimensional outputs that can be hard to interpret for both technical and non-technical experts \cite{an2020meeting}. To address this challenge we develop a visualisation strategy for the indicators; see Figure \ref{fig:model}. There we show the state- and specialty-specific resilience indicators that are described by glyphs of different shapes and colour. State-specific results include aggregated values of four indicators and are depicted by a diamond shaped glyph within a circle (a). The height and width of the diamond shape represent the states' resilience in terms of free capacity and lost patients, respectively. The colour-filled arcs within the semicircles describe the state level percentages of physicians that have higher than state-average risk and benefit scores. The physician- or specialty-specific resilience is described by two of the previous indicators, namely risk and benefit scores (c). These are visualised by the colour-coded halves of a heart-shaped symbol that can represent results of individual physicians as well as aggregated results of a medical specialty.

To illustrate how we measure these indicators in the network model, see Fig.~\ref{fig:model} (c,d). There we show the network neighbourhood of a physician ("physician 1") that becomes unavailable and has a relatively high risk score, meaning that his/her removal will affect many other physicians due to an inflow of new patients (c). As a result, physician 2 with a relatively low benefit score (meaning low capacity to accept new patients) becomes unavailable, and some patients need to contact yet another physician to find a new care provider and may potentially get lost (d). 


\newpage
\section{Data and Methods}
\subsection{Agent based model}

\subsection{Data-sets}
To construct the patient sharing network, we combine two data-sets: patient contacts for Austrian physicians as derived from administrative data, and a data set of physician's opening hours that was scraped from a health care information platform (www.herold.at) in March 2020\footnote{Permission for scraping the data was obtained from the owners of the platform}. Some physicians practice more than one medical specialty and/or practice in more than one municipality. Therefore, in the following we always refer to a unique (physician, specialty, municipality) combination, when talking about the number of physicians. A physician who is both, say, a general practitioner and an internist, or a physician who operates as a general practitioner in two municipalities is counted as two different physicians.

\subsection{Patient contacts}
The health record data provided by the Austrian Federal Ministry of Health contains all recorded patient visits (approx. 103,000,000) over a 1-year period from January 1, 2018 to December 31, 2018. The data contains patients' contacts to physicians from 13 different medical specialties and includes a pseudonymised unique ID for every patient and physician, the date of contact as well as the locations of physicians in form of 5-digit municipality IDs. The first digit of the municipality ID is further used to determine the federal state (9 states in total). The data set contains 9,580 unique combinations of (physician, specialisation, municipality).

\subsection{Patient sharing network}
To model connections between physicians, we use the patient contact data to construct a patient sharing network. In this network, every node is a physician. The network is described by the adjacency matrix $A_{i,j}$, where entries $a_{i,j}$ describe the number of shared patients between physicians $i$ and $j$, i.e. the number of patients that visited both physician $i$ and physician $j$ in a given period of time. The adjacency matrix is therefore symmetrical by definition and the network described by the adjacency matrix is undirected. To construct the adjacency matrix, we start with a patient's visit to physician $i$ and consider all visits of the given patient to other physicians during a period of three months before and after the initial visit to physician $i$. All thus identified physicians are therefore sharing one patient with $i$. This process is repeated for all patient visits of physician $i$. If there is at least one shared patient between $i$ and a given other physician $j$, there is an edge $a_{i,j}$ between $i$ and $j$. The edge is weighted by the sum of shared patients. This procedure is repeated for all physicians, resulting in a weighted undirected patient sharing network that is represented by the adjacency matrix $A_{i,j}$. 
To reduce spurious relations, the adjacency matrix can be thresholded by requiring a minimum number of shared patients and a maximum geographical distance between physicians $i$ and $j$. The distance is measured as the direct distance between $i$'s and $j$'s municipalities. The location of municipalities was determined as either the geolocation of the affiliated municipal office or as the geolocation of the area with highest population density in the municipality using a population density map. If the distance between physicians $i$ and $j$ exceeds threshold $d$, the edge is removed and $a_{i, j} = a_{j,i} = 0$. Similarly, when the number of shared patients between $i$ and $j$ is smaller than threshold $p$, the edge is removed.

To model the patient sharing network within a medical specialty, only the subset of nodes corresponding to physicians of the given specialty and their corresponding edges are used. Edges between physicians of different specialties are removed.

\subsection{Capacity estimation}
To estimate a physician's maximum capacity, $C$, we aggregate the patient contact data with information about opening hours; see SI Section "Matching of patient contact data to opening hours" for details. For robustness, we provide two different estimates of capacity: one based purely on the number of patient contacts ($C_p$) and one based on the opening hours and patient contacts ($C_h$). 

For the patient-contact-based capacity estimate $C_p$, we calculate the median patient capacity (number of quarterly patient contacts) of the top 10\% of physicians for each specialty. We then assign this median capacity to the rest of the physicians (not top 10\% capacity), regardless of opening hours. Since this approach does not differentiate between physicians with different opening hours, it is prone to overestimating their capacity.

For the opening-hour-based capacity estimate $C_h$, we divide physicians of every specialty into bins based on their total number of weekly opening hours with a bin size of 5 opening hours. For each bin, we then identify the top 10\% of physicians in terms of their capacity (number of quarterly patient contacts) and calculate their median capacity. This median capacity is then assigned to all remaining physicians in the bin as their maximum patient capacity.

By merging the information of the two data-sets we get a comprehensive description of the majority of physicians in the country. We can initially describe each physician with specific characteristics: the specialty, municipality, number of patients seen per quarter ($N$) and total capacity based on opening-hours ($C_h$). While the specialty and location are constant, the current number of patients and the remaining free capacity change during simulations. General descriptive information is shown in Tab.~\ref{tab:baseline}. The vast majority of physicians are general practitioners ($n=4,967$), while psychiatrists ($n=161$) form the smallest specialty group. Patient variables in the model include their starting location (municipality) based on the physician they start at, the number of displacement steps they have already undergone, step-wise travelled distance and total travelled distance.

\begin{table}
    \centering
    \resizebox{\columnwidth}{!}{\begin{tabular}{l|r|r|r|r|r}
    specialty & Abbrev. & Physicians  & Opening hours  &  Patients per quarter & Maximum capacity  \\
    \toprule
    General medicine & GP & 4967 & 19 ($\pm$6) & 2949 ($\pm$2246)  &  6391 ($\pm$1825) \\
    Ophthalmology & OPH  & 436 & 21 ($\pm$5)   & 2234 ($\pm$1955) & 4498 ($\pm$2222)  \\
    Surgery &  SRG & 278 & 25 ($\pm$11)  & 1874 ($\pm$3025)  &  6495 ($\pm$3870) \\
    Dermatology & DER & 290  &  21 ($\pm$4)   & 2660 ($\pm$2179)  & 5634 ($\pm$1722)  \\
    Gynaecology and obstetrics & OBGYN & 552  & 22 ($\pm$4)   & 1357 ($\pm$1240)  &  2887 ($\pm$1283) \\
    Otorhinolaryngology & ENT & 293  & 22 ($\pm$6)   &  1927 ($\pm$1235) & 4019 ($\pm$769)  \\
    Internal medicine & IM & 1058  & 22 ($\pm$6)   & 1601 ($\pm$3424)  &  4900 ($\pm$5595) \\
    Paediatrics & PED & 336  & 21 ($\pm$6) & 2212 ($\pm$1788)  &  4835 ($\pm$3156) \\
    Neurology & NEU & 166 & 22 ($\pm$5)  & 1279 ($\pm$1173)  & 2681 ($\pm$1596)  \\
    Orthopaedics & ORTH  & 366  &  22 ($\pm$7)  & 4150 ($\pm$6091)  &  13522 ($\pm$5621) \\
    Psychiatry & PSY &  161 & 21 ($\pm$5)  & 1056 $\pm$1115)  &  2776 ($\pm$1027) \\
    Radiology & RAD &  424 & 42 ($\pm$10)  & 3759 ($\pm$4283)  &  12093 ($\pm$4481) \\
    Urology & URO & 253  & 20 ($\pm$4)  & 1558 ($\pm$1014)  &  3161 ($\pm$581)  \\
    \bottomrule
    \end{tabular}}
    \caption{Descriptive table of medical specialties and their basic characteristics. Physicians in the data-sets were divided into one out of 13 specialties. Numbers show mean and standard deviation. Maximum capacities $C_h$ were calculated based on opening hour information, using the top 10\% of physicians per specialty.}
    \label{tab:baseline}
\end{table}

\subsection{Model parameters and robustness}
The model contains a number of parameters that can be varied and tested for robustness. We define a baseline parameter setting as follows: The maximum number of displacement steps before a patient is considered lost ($s = 10$), the median patient capacity of top c\% of most visited physicians of certain specialty that is assigned to all other physicians as their maximum capacity ($c = 10$), the maximum distance in kilometres patients are willing to travel to a new physician ($d = 100$), the number of times the simulation attempts to find a physician within this distance from the starting location of a patient before choosing a physician located further away ($i = 10$), the minimum number of shared patients in the adjacency matrix for a valid connection ($p = 2$) and the probability $\alpha$ of random re-locations of patients to unconnected physicians (which in the baseline setting does not occur, $\alpha = 0$). Physicians who are not connected to the giant component~\cite{bollobas2001evolution} of the patient sharing network remain in the system for the possibility of patients to choose them at random (only if $\alpha > 0$).


\subsection{Free capacity \& lost patients}
The initial free capacity in each federal state is reduced in each simulation step as more and more physicians become unavailable. We track how the remaining relative free capacity in each state is filled up/diminished until there is no capacity left and all the physicians are removed from the system. We define a critical limit of 20\% of the remaining free capacity per state and track the average relative number of physicians that needs to be removed to reach this critical free capacity limit, $L_{FC}$. 

Similarly, we sum the number of lost patients and calculate the average relative number of cumulative lost patients in each federal state. The more physicians are unavailable, the lower the remaining free capacity and the higher the probability for a patient to get rejected at a new physician - after all physicians are removed, 100\% of patients are lost. Analogous to the free capacity we define a critical lost patient limit of 1\% and track the relative number of physicians removed before this point is reached, $L_{LP}$.

\subsection{Risk \& benefit scores}
Based on the initial physicians' characteristics we can describe the negative and positive influence that physician contributes to the healthcare system by defining a risk and benefit score for each physician. The risk score $R_i$ represents physician $i$'s risk and is measured by the average extra load of patients that their first-degree neighbours in the patient sharing network must bear in case of $i$'s unavailability. We therefore define the risk score as
\begin{align*}
    R_i = \left< \mathrm{min}\left(\frac{N_j + N_i \cdot w_j}{C_j}, 1\right) \right>_j\,,
\end{align*}
where $N_i$ and $N_j$ stand for the number of patients of physician $i$ and their neighbours $j$, $C_j$ is $j$'s total capacity and $w_j$ describes the normalised connection weight between $j$ and $i$. The range of the risk score is $[0,1]$, where 0 corresponds to the lowest and 1 to the highest risk.
The benefit score $B_i$ represents physician $i$'s beneficial contribution to the healthcare system in terms of their initial free capacity. The benefit sores are normalised over all physicians of a given medical specialty and range as well $[0,1]$, with 0 being the lowest benefit and 1 the highest benefit.

\section{Results}

\subsection{Regional resilience indicators}
\begin{figure*}
    \centering
    \includegraphics[width=1\textwidth]{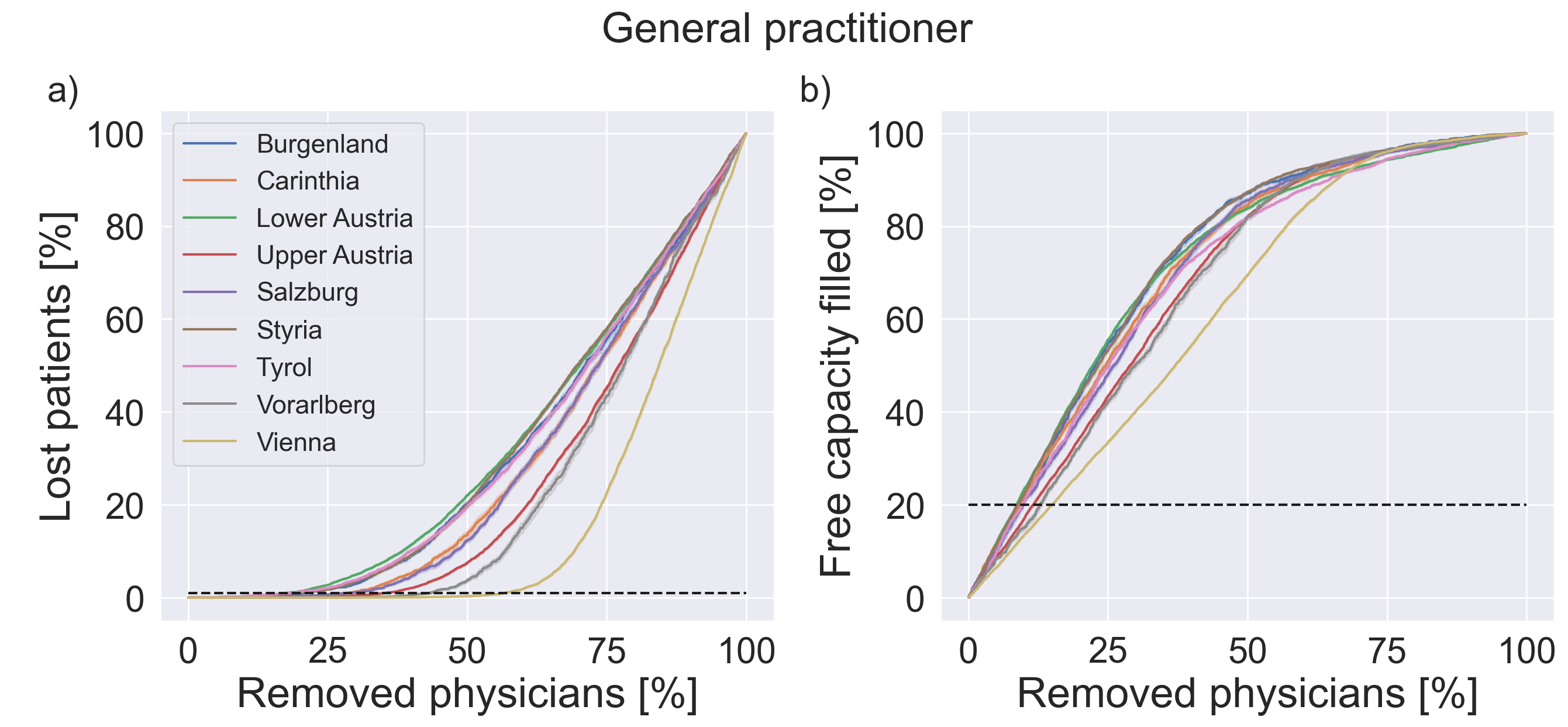}
    \caption{Development of lost patients (a) and free capacity (b) in federal states. Averaged model results for the example of general practitioners as physicians are gradually removed from the system. Dashed lines indicate critical limits of 1\% lost patients and 20\% remaining free capacity, respectively.}
    \label{fig:GP}
\end{figure*}

To investigate indicators of regional resilience of the healthcare system, we report the development of the relative lost patients as well as the remaining free capacity as physicians are gradually removed in the simulation. We report averaged results over 100 simulation iterations. Figure \ref{fig:GP} (a) shows (for the example of general practitioners [GP]) the development of relative lost patients and free capacity during the course of the simulation for all federal states.
Initially, the removal of the first few physicians can be absorbed by the system without losing patients. At a specific percentage of removed physicians, patients start becoming "lost", e.g. unable to find another physician. The point at which the number of lost patients surpasses 1\% varies widely between different states, ranging from around 20\% of removed physicians (Lower Austria, Tyrol) to more than 50\% (Vorarlberg, Vienna).

Figure \ref{fig:GP} also shows how fast the initial free capacity is filled up in each federal state as GPs are gradually removed from the system. Under the sequential removal of physicians, the remaining free capacity first decreases proportional to the number of physicians that were removed removed. The slopes of this linear dependence, however, vary widely across federal states and relate to the lost patients indicator: the faster capacity decreases, the sooner patients get lost. As the system gradually approaches a state with no remaining capacity, there is no more free capacity to accommodate displaced patients and an increasing number of patients get lost. 

\begin{figure*}
    \centering
    \includegraphics[width=1\textwidth]{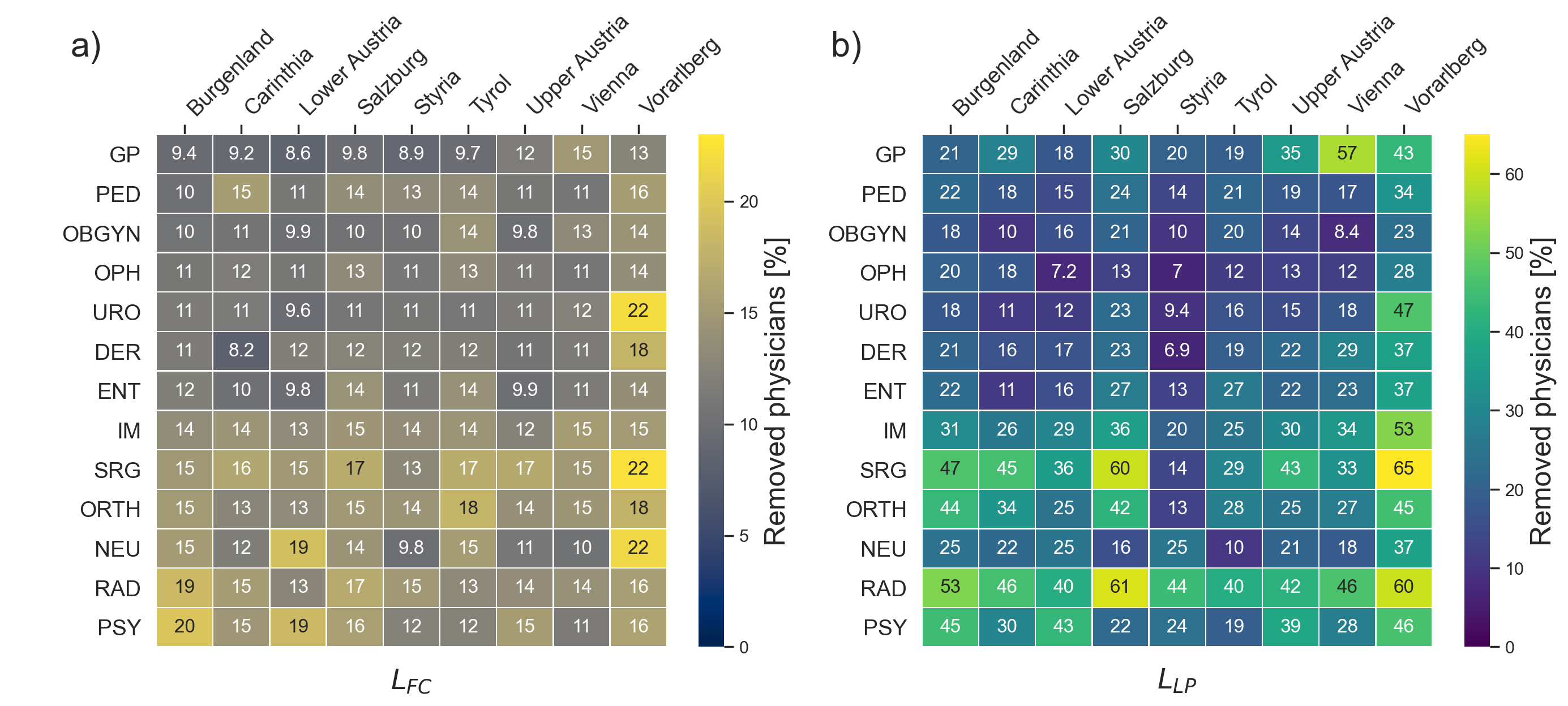}
    \caption{Average critical \% of removals in federal states per specialty until limits of free capacity (a) and lost patients (b) are reached. Critical limits are 1\% for lost patients and 20\% for remaining free capacity. Darker colours indicate higher number of physicians that can be removed before exceeding critical limits. Tables including the standard deviation of simulation results are given in Tables S2 and S3.}
    \label{fig:heatmap}
\end{figure*}

\begin{figure*}[ht]
    \centering
    \includegraphics[width=1\textwidth]{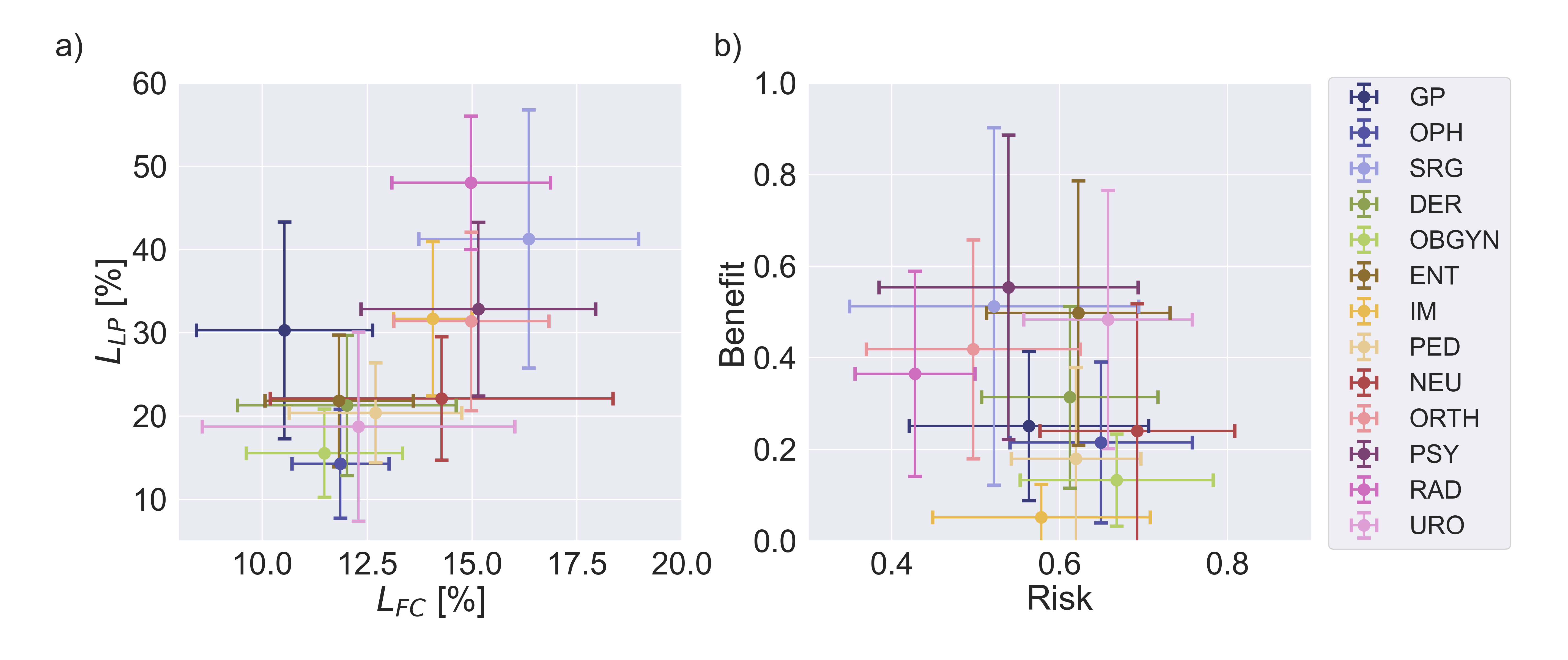}
    \caption{Country-level specialty specific simulation results. a) shows specialty specific risk and benefit scores averaged over all states, b) shows state averaged percentages of physicians that can be removed before critical limits for lost patients and free capacity are reached.}
    \label{fig:Results}
\end{figure*}

In the SI, Figures S4 to S15, 
 we show lost patients and free capacity as a function of the percentage of removed physicians for all other specialties. A summary for all states and specialties and the relative number of physicians that can be removed from the system until critical limits are reached (20\% for the remaining free capacity and 1\% for lost patients) is shown in Fig.~\ref{fig:heatmap}. The resilience indicators vary widely across federal states and specialties. Resilience indicators in the state of Styria are typically on the lower side, whereas structures in Vorarlberg appear to be substantially more resilient. Psychiatrists, neurologists, surgery, internal medicine, orthopaedics and radiology show a tendency to more resilient care networks compared to ophtalmology, dermatology, or urology. This can also be seen in Fig.~\ref{fig:Results}, where we show nationwide averages of these resilience indicators.

In addition to regional resilience of the healthcare system as a whole, we also investigate the individual risk and benefit a given physician contributes to the system. To this end, for each specialty $j$ we average the individual risk($\mathrm{risk} _j$) and benefit ($\mathrm{benefit} _j$) scores over all federal states to obtain nationwide results. Figure~\ref{fig:Results} (b) shows the mean and standard deviation for each specialty averaged over all federal states.

In general, there is a tendency that specialties with a high risk score and/or a low benefit score are also less resilient in terms of lost patients and free capacity. To quantify this observation, we evaluate a linear regression model of the form $ L_{FC_j} \; (L_{LP_j}) \sim r_j \cdot  \mathrm{risk} _j + b_j \cdot  \mathrm{benefit} _j + \mathrm{const.}$,
where $L_{FC_j}$ and $L_{LP_j}$ are the state-averaged percentages of physicians of specialty $j$ removed before reaching critical limits of free capacity and lost patients, respectively. The averaged effect sizes and standard deviations of coefficients $r$ and $b$ over all $j$ confirm the correlation between these indicators as $\langle r \rangle = -11 (\pm 10)$, $\langle b \rangle = 25 (\pm 28)$ for $L_{FC}$ and $\langle r \rangle = -70 (\pm 35)$, $\langle b \rangle = 58 (\pm 86)$ for $L_{LP}$. However, there is no discernible correlation between the risk and benefit scores across the specialties. This suggests that these scores capture two structurally independent properties that together determine the resilience of a specialty. This also motivates the use of the individual-level risk and benefit scores to assess the contributions of physicians to the systemic resilience of these regional care sectors.

\subsection{Interactive online visualisation tool}
To visualise the simulation results for the resilience indicators, we develop a visualisation strategy that can be found at \href{https://vis.csh.ac.at/care-network-resilience/}{https://vis.csh.ac.at/care-network-resilience/}. Using the aggregated simulation data and the network structure of physicians within the healthcare system, we show detailed state- and specialty-specific information on resilience indicators, as well as number of physicians and patients. Additionally, the tool allows for the possibility of interactively removing physicians from the system whilst updating the resilience values. Exemplary results for psychiatrists in all Austrian states are shown in panel (e) of Fig. ~\ref{fig:model}.

\subsection{Robustness}
We performed several robustness tests to assess the sensitivity of our results. We find that varying the maximum capacity parameter $c$ ($c=20, 30, 40$) introduced most variation in the results. However, this variation does not change the regional resilience indicators qualitatively, meaning that the specific values for overstepping critical limits of lost patients or free capacity will differ for states when using different $c$, but it will not convert a "more resilient" state to a "less resilient" state compared to others (ranking remains robust). For the purpose of this stress-test we selected $c=10$ as it necessitates a large proportion of physicians to increase their available capacities.

Increasing the probability to approach physicians at random instead of following the links in the patient-sharing network, i.e. setting $\alpha=0.15$, showed the second largest variation in simulation outcomes. Again, rankings remained robust as increasing $\alpha$ gradually washes out the network effects. In the absence of such network effects (for $\alpha=1$) our resilience indicators would be determined by physician density alone. Other simulation parameters only show minor changes in simulation results. Every other model parameter than $c$ and $\alpha$ was tested with at least one additional setting ($s=5, d=50, i=5, p=1$) to assess robustness of the results.

\section{Discussion}

Fast progress in digital technology stimulates debates on new consultation methods. Especially the Covid-19 pandemic encouraged and increased the implementation of online consultations in many fields of medicine \cite{bashshur2020telemedicine, mann2020covid}. However, patients in regions with more deprived populations appear to be less likely to benefit from telehealth \cite{chunara2021telemedicine}. Patients with inadequate digital devices or in regions with poor network connectivity will have problems to access telehealth, possibly even widening already existing treatment gaps. On the other hand, remote GP but also specialist consultations may be helpful and convenient for many patients with chronic diseases in need of regular follow-up visits. However, a certain number of face-to-face consultations will remain necessary to provide adequate patients´ care in the future. Therefore, although telehealth may help to diminish problems of physicians’ shortage in some areas in the future, allowing continued access to services, we will still need a sufficient number of experienced physicians with capacities to perform online consultations in addition to standard clinical care. 

Experiences during the Omicron-variant-related surge in the SARS-CoV-2 pandemic have demonstrated that scenarios in which ten to twenty percent of the health workforce are absent over several weeks can indeed happen~\cite{Torjesen2021}.
As health care demand is further expected to increase in developed countries as the population is ageing and the prevalence of chronic conditions is increasing~\cite{Beard15}, there is an increasing need to know how far the available physician supply can be stretched to still deliver the desired quality of care.
In this work we proposed novel regional resilience indicators to quantify the point beyond which further decreases in physician densities would begin to severely impair the population's ability to access primary or secondary care. 
We find strong heterogeneities in these resilience indicators across different medical fields and regions. Our findings show that the naturally emerging patient sharing networks that can be reconstructed from routinely collected administrative data are a valuable source of information when assessing the resilience of the healthcare system.

Our resilience indicators can be disaggregated into indicators for individual physicians to quantify the extent to which their hypothetical removal would stress the system (risk score) or how much of the stress from the removal of other physicians they would be able to absorb (benefit score).
These indicators allow health authorities to rapidly identify physicians with sufficient capacities to take over supply and alert them that they might receive larger than expected inflows of patients in the situation where other physicians retire.
This could help planners to proactively anticipate whether such retirements could cause bottlenecks in certain regions and whether the now vacant position should be replaced with a higher level of priority.
To highlight this potential of our approach, we have developed an interactive visualisation framework that allows health care planners and professionals to browse and intuitively access the regional and individual resilience indicators as well as to inspect these naturally emerging physician networks and how potential absences would impact them.

Our study has several limitations. Firstly, we focused on patient-sharing networks among physicians of only the same medical specialty. However, especially in rural areas GPs are used to cover and treat a broad spectrum of diseases and they may be able to attenuate the gap caused by shortage of some specialists \cite{selby2018primary}. On the other hand, specialists in internal medicine could temporarily replace GPs in some areas. Moreover, in Austria many tasks of primary care are still transferred or covered by outpatient departments of bigger hospitals in the cities. Therefore, shortfall of physicians might have different effects in rural and urban areas.

 Secondly, the results of the agent-based simulation depend on the assumptions regarding various model parameters. We performed several robustness tests to better understand the sensitivity of our results. We find that our results are most sensitive to changes in the maximum capacity parameter $c$ and teleportation parameter $\alpha$. However, we observed no substantial changes in the rankings of states and specialties, respectively. 

Finally, the sex of physicians should be included in future analyses because recent trends show that the number of female physicians is steadily increasing but they appear to have different demands and preferences regarding work-life balance, working models and fields of activities and duties \cite{pelley2020specialty}.  Moreover, they need specific infrastructures like family-friendly workplace. 

Our work demonstrates that the resilience of primary and specialist care is an emergent property of formal and informal networks that physicians form amongst themselves.
While fiscal constraints and concerns regarding an ageing physician workforce in rural areas are growing worldwide \cite{skinner2019implications}, we find that these care networks indeed possess tipping points in terms of their ability to provide care to the entire population.
Removing or losing an alleged excess capacity or oversupply of physicians might inadvertently push healthcare systems closer to their tipping points.
Using our indicators allows health authorities to quantify how close they are to these tipping points and thereby strike a more data-informed balance between system resilience and effectiveness.

\section{Acknowledgments} 

MK, AKW, and PK acknowledge financial support from the Medizinisch Wissenschaftlicher Fonds des Buergermeisters der Bundeshauptstadt Wien under CoVid004. J.S. and S.T. acknowledge financial support from the Austrian Science Promotion Agency FFG under 882184.

\section{Author Contributions}
M.K., J.L., S.T., and P.K. designed research; M.K., J.L. and R.L.S. performed
research; J.S. and L.Y. contributed analytic tools; M.K., J.L., R.L.S., J.S. and L.Y. analyzed data; All authors reviewed and contributed to the manuscript.


\pagebreak

\appendix

{\Large Supporting Information}

\renewcommand\thefigure{S\arabic{figure}}    
\setcounter{figure}{0}    

\renewcommand\thetable{S\arabic{table}}    
\setcounter{table}{0} 

\section{Matching of patient contact data to opening hours}
To assess the patient contact capacity of physicians, we obtained the permission to scrape data about opening hours from the platform \url{www.herold.at} for Austrian physicians in March 2020. The scraped data includes a physician's unique ID, speciality, municipality, address, information about whether s/he is a panel physician and opening times for every day of the week. The opening hour data includes a total of 49,562 unique (physician, specialisation, municipality) combinations, 

Before probabilistically matching the opening hour data to the patient contact data, we perform a number of cleaning and data enrichment steps: We only include panel physicians in our analysis, reducing the number of physicians to 7,831. We further exclude all physicians that do not indicate any opening hours (1,137 physicians). Physicians that indicate opening hours only "by appointment" on a given day are assumed to be open for 2 hours on that day. If there is more than one entry of the same physician but with different opening hours (29 entries in total), we keep the entry with the longer opening hours. We then match the 29 specialities given in the opening hours data set (see table S1) to the 13 specialities given in the patient contact data (see Table 1). We drop 9 specialities – such as medical and chemical laboratory diagnostics – that are not corresponding to any of the specialities in the patient contact data (90 physicians). This leaves us with a total of 6,604 unique physicians.

We then probabilistically assign the opening hour data to the patient contact data following a two-step process: We first look for all instances where there is only a single physician with a given speciality in a given district in both data sets and match these physicians. This results in 1,233 direct matches. For the remaining physicians, we calculate the vector of proportional opening times $\mathbf{v}$ for every day of the week: if a physician has a total of $h=32$ opening hours of which 8 fall on Mondays, 8 on Tuesdays, 8 on Wednesdays, 4 on Thursdays and 4 on Fridays, this results in an opening hour vector of $\mathbf{v} = (8/h, 8/h, 8/h, 4/h, 4/h, 0/h, 0/h) = (0.25, 0.25, 0.25, 0.125, 0.125, 0, 0)$. We also calculate the vector of proportional patient contacts $\mathbf{w}$ for the physicians in the patient contact data set: if a physician has a total of $n=10,000$ patient contacts throughout the year, of which 1,500 occurred on Mondays, 1,500 on Tuesdays, 3,000 on Wednesdays, 2,000 on Thursdays and 2,000 on Fridays, this results in a patient contact vector of $\mathbf{w} = (1500/n, 1500/n, 3000/n, 2000/n, 2000/n, 0/n, 0/n) = (0.15, 0.15, 0.3, 0.2, 0.2, 0, 0)$. 

We then assign physicians with a given speciality and within a given municipality based on the difference between their opening hours and patient contacts by minimizing $E = \sum_{i=1}^7 \left| v_i - w_i \right|$. We also introduce a threshold $\epsilon = 0.5$ and only match two physicians if $E \leq \epsilon$. This results in 2,891 additional matches (2,537 matches if $\epsilon = 0.3$ and 3,055 matches if $\epsilon=0.7$). At the end of the procedure, we successfully assigned capacities to 4,288 physicians from the opening hour data set to the patient contact data set, leaving 2,406 physicians (35.9\%) in the opening hour data set and 5,292 physicians (55.2\%) in the patient contact data set unmatched. We perform a left join of the physicians from the patient contact data and the matched opening hour information, resulting in a total of 9,580 physicians of which 44.8\% have opening hour information.  

Inspection of the distribution of opening hours of matched and unmatched physicians (see SI) shows that there is a slight bias towards unmatched physicians having fewer opening hours. This difference is significant ($p < 0.05$, two-sided t-test of the opening hour distributions) for general practitioners ($\bar{h}_\mathrm{matched} = 20.3\pm 4.7\;$h, $\bar{h}_\mathrm{unmatched} = 19.7\pm6.8\;$h, mean $\pm$ standard deviation), urologists ($\bar{h}_\mathrm{matched} = 20.5\pm5.0\;$h, $\bar{h}_\mathrm{unmatched} = 18.6\pm7.4\;$h) and dermatologists ($\bar{h}_\mathrm{matched} = 21.4\pm5.1\;$h, $\bar{h}_\mathrm{unmatched} = 19.6\pm5.0\;$h). We see a similar trend in the distributions of patient contacts (see SI), as unmatched physicians tend to have fewer yearly patient contacts. This difference is significant ($p<0.05$) for general practitioners ($\bar{n}_\mathrm{matched} = 16054\pm7881\;$h, $\bar{n}_\mathrm{unmatched} = 10770\pm6688\;$h), internists ($\bar{n}_\mathrm{matched} = 15032\pm26184\;$h, $\bar{n}_\mathrm{unmatched} = 7165\pm6201\;$h), gynaecologists ($\bar{n}_\mathrm{matched} = 7125\pm6579\;$h, $\bar{n}_\mathrm{unmatched} = 5608\pm3723\;$h), ophtalmologists ($\bar{n}_\mathrm{matched} = 11330\pm10254\;$h, $\bar{n}_\mathrm{unmatched} = 8174\pm5767\;$h), otolaryngologists ($\bar{n}_\mathrm{matched} = 8925\pm4903\;$h, $\bar{n}_\mathrm{unmatched} = 7667\pm4463\;$h), dermatologists ($\bar{n}_\mathrm{matched} = 13582\pm11496\;$h, $\bar{n}_\mathrm{unmatched} = 9891\pm5996\;$h), orthopaedists ($\bar{n}_\mathrm{matched} = 24376\pm33647\;$h, $\bar{n}_\mathrm{unmatched} = 13062\pm14792\;$h) and radiologists ($\bar{n}_\mathrm{matched} = 25616\pm27148\;$h, $\bar{n}_\mathrm{unmatched} = 13799\pm10436\;$h). The matching rate (number of matched physicians / total number of physicians) is best for general practitioners (55.1\%) and worst for internists (20.7\%) (see SI).

For the unmatched physicians, we impute the opening hour information in the following way: for every unmatched physician, we look for all successfully matched physicians with the same specialisation and similar number of patients. We first look for all physicians that have the same number of annual patients $\pm 10$. If no physicians are found, we increase the search interval to $\pm 100$ or $\pm 1000$ annual patients. We then calculate the average opening hours for each day of the week for these successfully matched physicians and use this information to impute the missing opening hour information for the unmatched physicians.

\begingroup
\setlength{\tabcolsep}{10pt} 
\renewcommand{\arraystretch}{0.55} 
\begin{table}
    \centering
    \footnotesize{
    \begin{tabular}{p{7cm}|p{7cm}|p{1cm}}
         opening hours & patient contacts & n \\
         \midrule
         general medicine & general practitioner & 3552 \\
         internal medicine & internist & 423 \\
         obstetrics and gynaecology & gynaecologist & 395 \\
         ophtalmology and optometry & ophtalmology & 363 \\
         paediatric medicine & paediatrician & 258 \\
         otorhinolaryngology, ophthalmology and angiology & otorhinolaryngologist, ophthalmologist and angiologist & 238 \\
         orthopaedics and orthopaedic surgery & orthopaedist & 232 \\
         dermatology and sexually transmitted diseases & dermatologist & 226 \\
         radiology & radiologist & 213 \\
         urology & urologist & 178 \\
         psychiatry & psychiatrist & 174 \\
         pulmonology and pneumology & internist & 148 \\
         surgery & surgeon & 118 \\
         neurology & neurologist & 67 \\
         medical and chemical laboratory diagnostics & – & 32 \\
         childhood and adolescent psychiatry & – & 19 \\
         physical medicine and general rehabilitation & – & 17 \\
         trauma surgery & surgeon & 17 \\
         clinical pathology and molecular pathology & – & 8 \\
         clinical microbiology and hygienics & – & 5 \\
         dental, oral and maxillo-facial surgery & – & 3 \\
         plastic, reconstructive, and aesthetic surgery & – & 3 \\
         childhood and adolescent psychiatry and psychotherapy & – & 2 \\
         neurosurgery & neurosurgeon & 2 \\
         anesthesiology and intensive care medicine & – & 1 \\
         \bottomrule
    \end{tabular}
    }
    \caption{Mapping of physician's specialities between the opening hour data set and the patient contact data set.}
    \label{tab:speciality_mapping}
\end{table}
\endgroup

\begin{figure}
    \centering
    \includegraphics[width=0.9\textwidth]{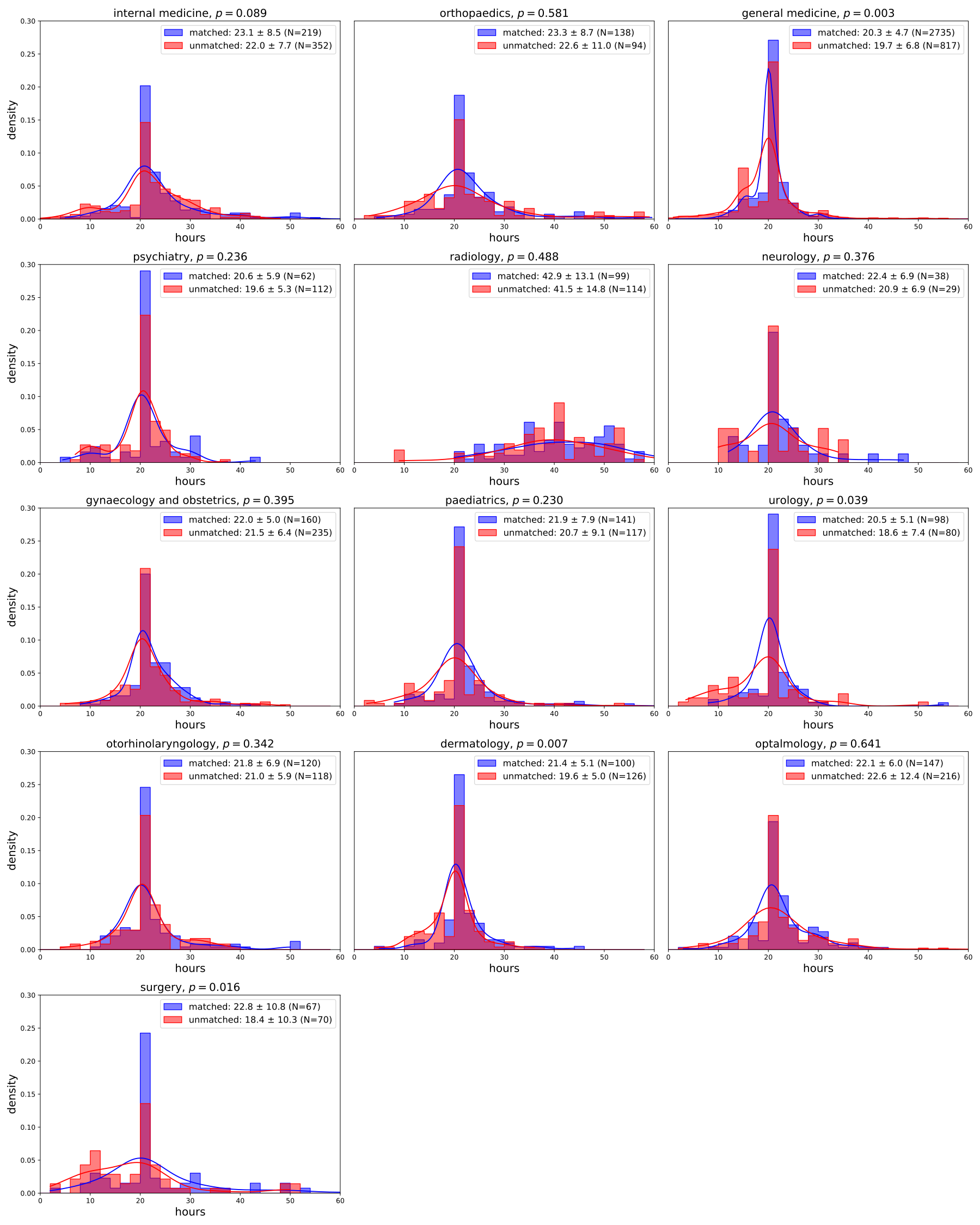}
    \label{fig:opening_hour_bias}
    \caption{Distribution of opening hours of matched (blue) and unmatched (red) physicians from the opening hour data set. P-values from two-sided independent t-tests between the distributions are indicated for each speciality.}
\end{figure}

\begin{figure}
    \centering
    \includegraphics[width=0.9\textwidth]{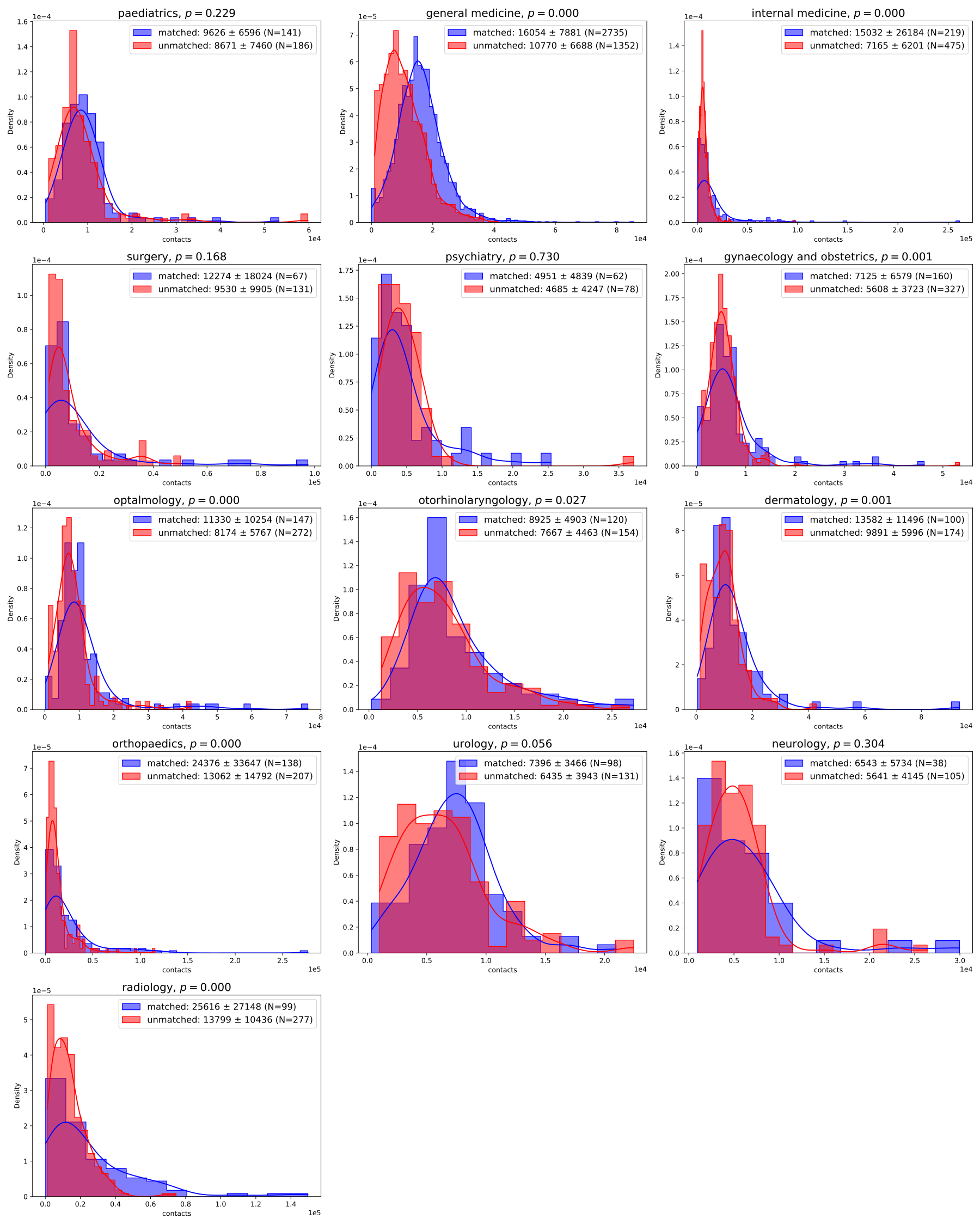}
    \label{fig:patient_contact_bias}
    \caption{Distribution of patient contacts of matched (blue) and unmatched (red) physicians from the patient contact data set. P-values from two-sided independent t-tests between the distributions are indicated for each speciality.}
\end{figure}

\begin{figure}
    \centering
    \includegraphics[width=\textwidth]{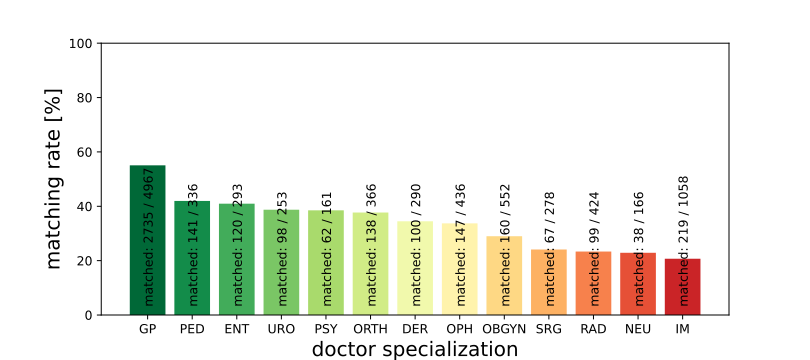}
    \label{fig:matching_rates_by_specialization}
    \caption{Matching rates for in \% matched physicians from the patient contact data set for each of the 13 specialities.}
\end{figure}

\newpage
\section{Progress of resilience indicators in states}
The following figures (from Figure ~\ref{fig:OPH} to Figure ~\ref{fig:URO}) show state-level simulation results for the continuous physician removal process for different specialties (except GP). Parts a) show the cumulative relative lost patients in states, parts b) show gradually filled up free capacity in states.

\begin{figure}
    \centering
    \includegraphics[width=0.9\textwidth]{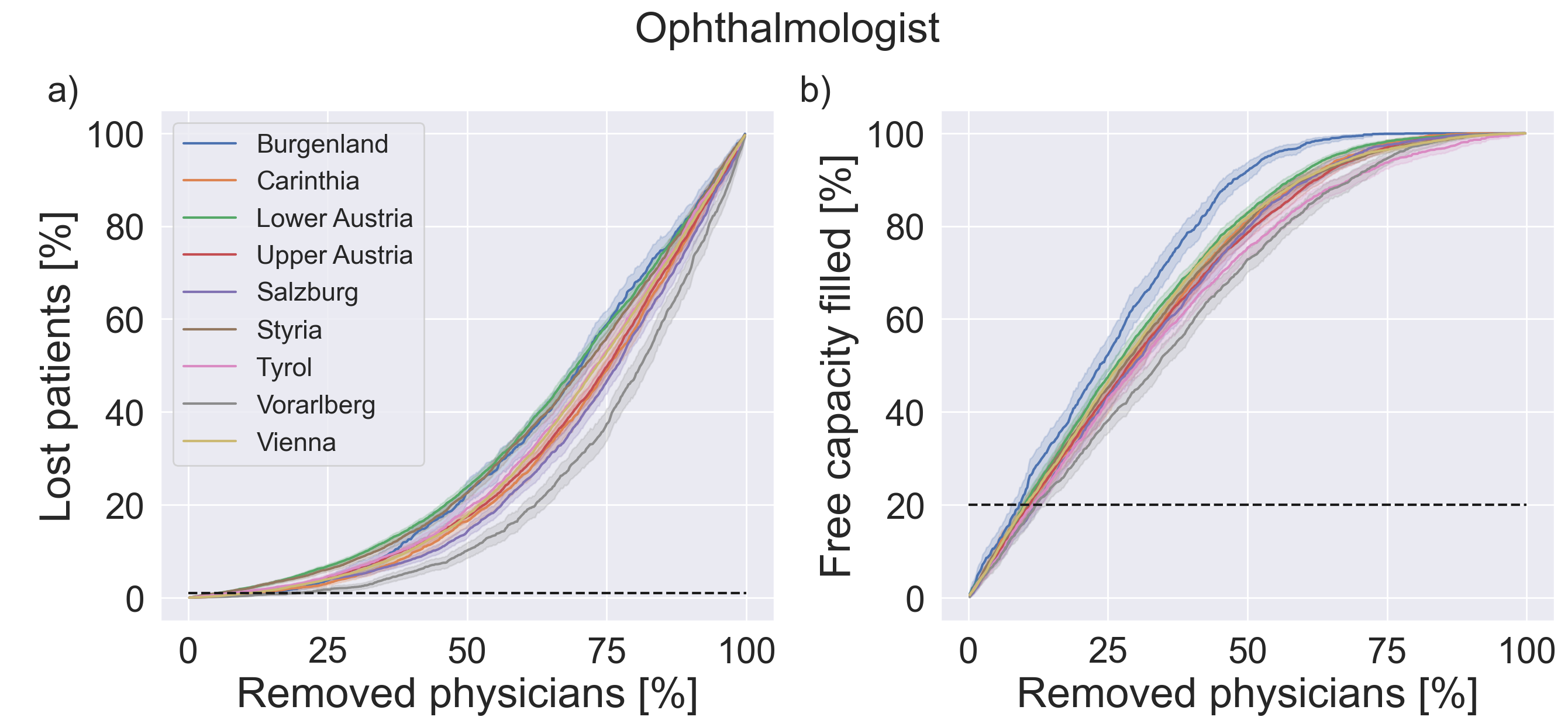}
    \caption{Lost patients and free capacity as a function of removed physicians for ophthalmologists.}
    \label{fig:OPH}
\end{figure}

\begin{figure}
    \centering
    \includegraphics[width=0.9\textwidth]{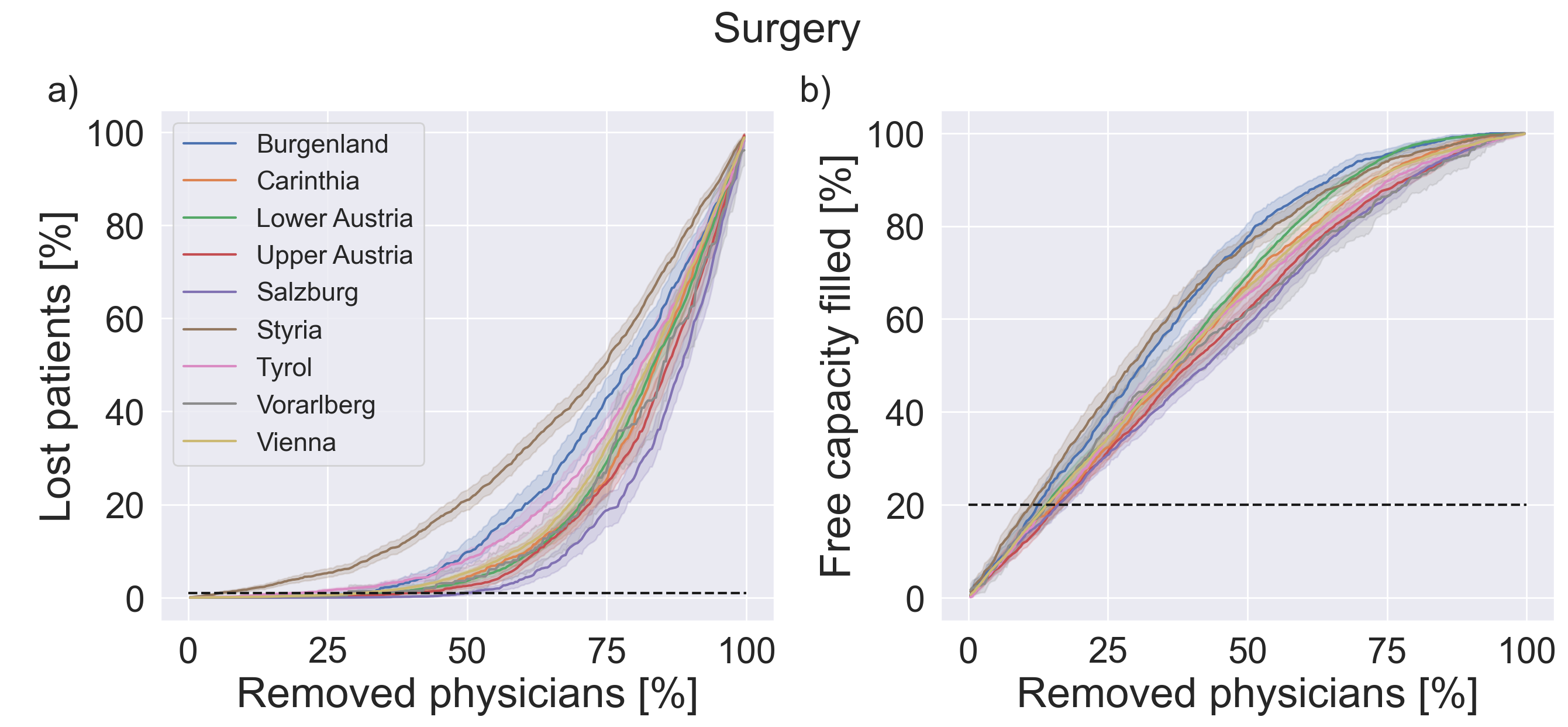}
     \caption{Lost patients and free capacity as a function of removed physicians for surgery.}
    \label{fig:SRG}
\end{figure}

\begin{figure}
    \centering
    \includegraphics[width=0.9\textwidth]{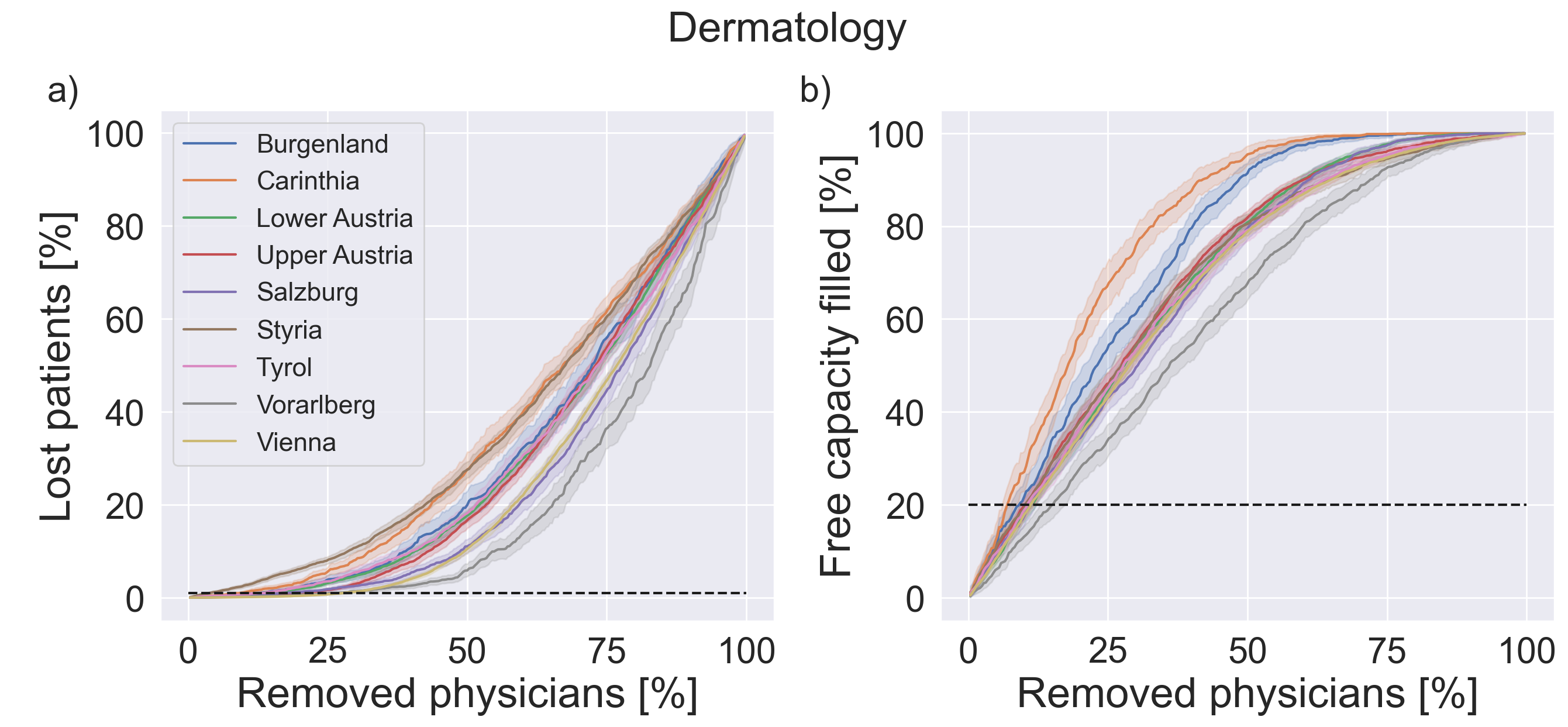}
     \caption{Lost patients and free capacity as a function of removed physicians for dermatologists.}
    \label{fig:DER}
\end{figure}

\begin{figure}
    \centering
    \includegraphics[width=0.9\textwidth]{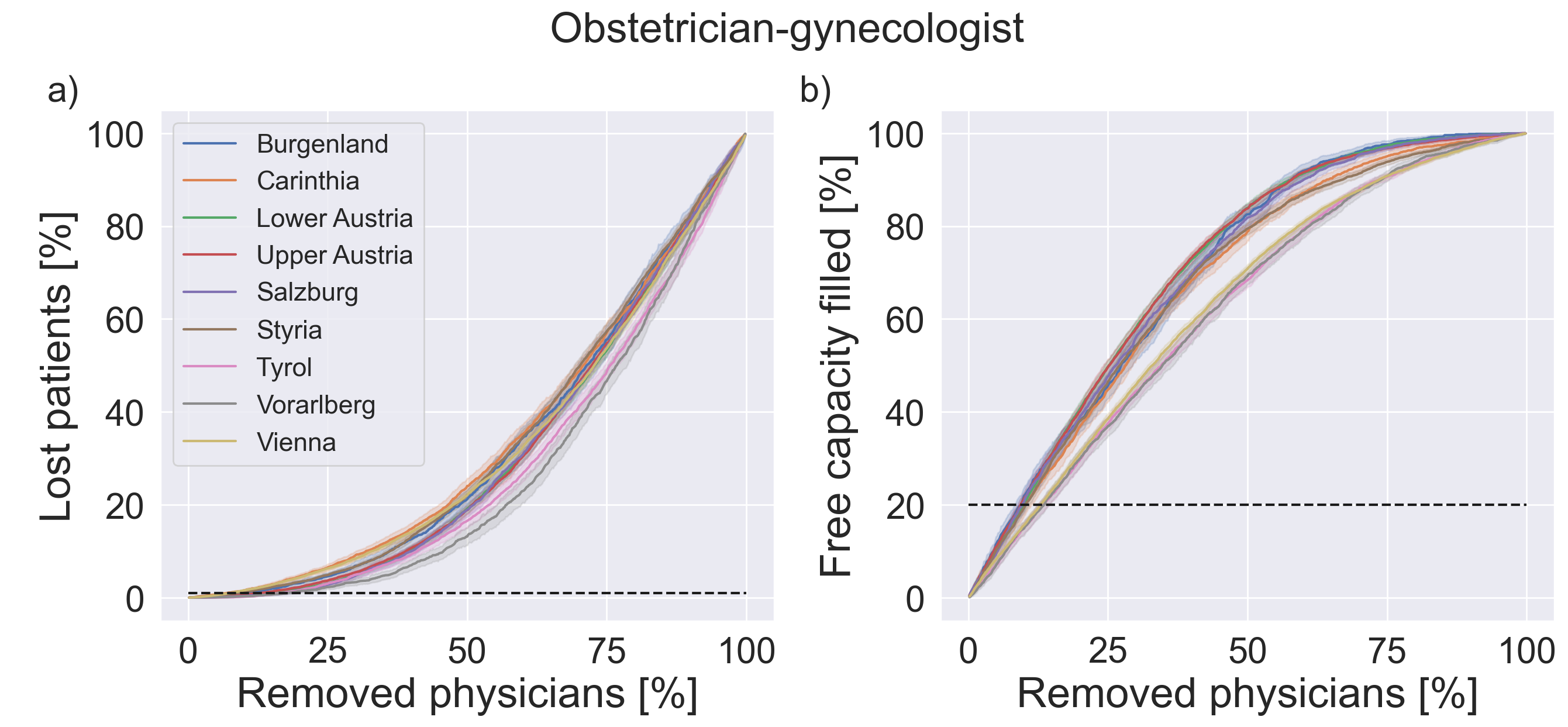}
     \caption{Lost patients and free capacity as a function of removed physicians for obstetricians/gynecologists.}
    \label{fig:OBGYN}
\end{figure}

\begin{figure}
    \centering
    \includegraphics[width=0.9\textwidth]{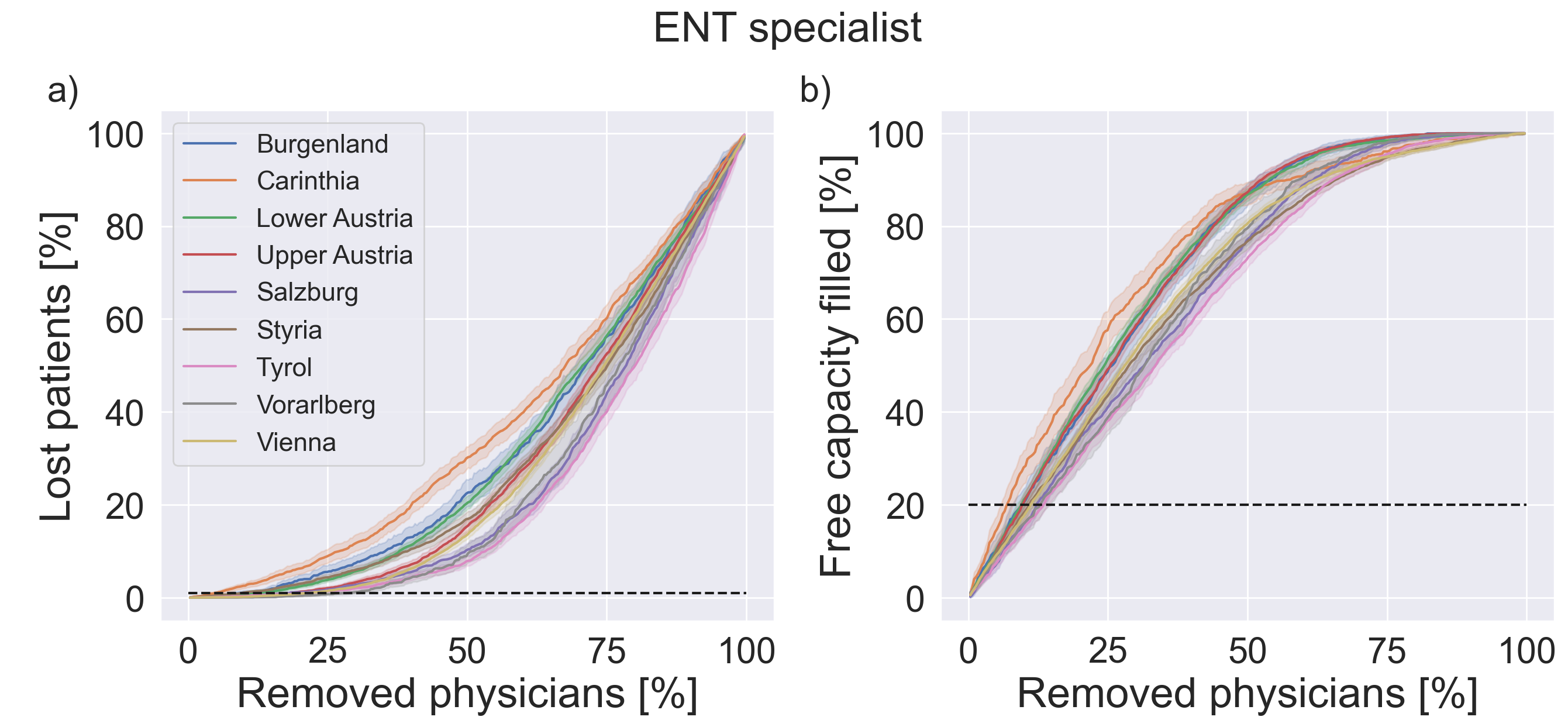}
     \caption{Lost patients and free capacity as a function of removed physicians for ENT specialists.}
    \label{fig:ENT}
\end{figure}

\begin{figure}
    \centering
    \includegraphics[width=0.9\textwidth]{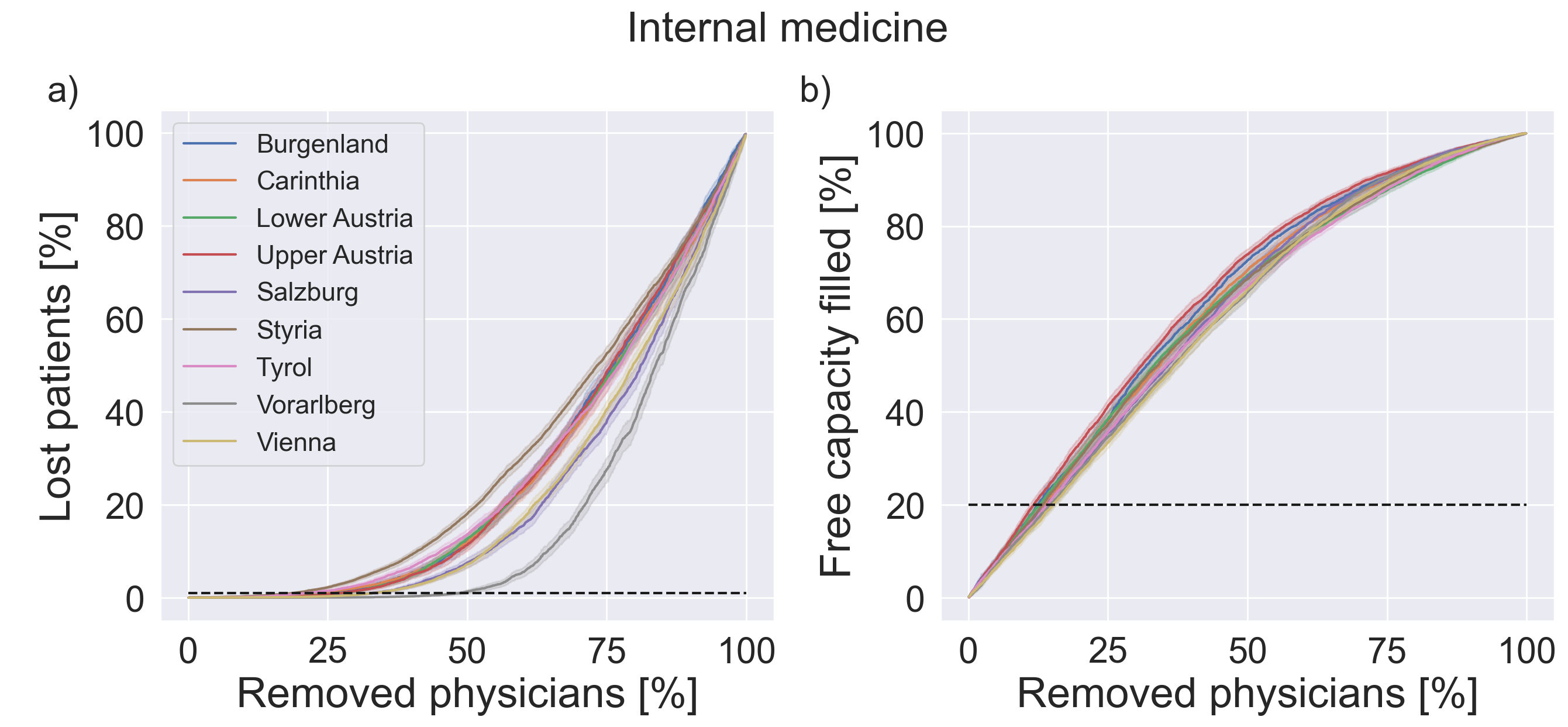}
     \caption{Lost patients and free capacity as a function of removed physicians for internal medicine.}
    \label{fig:IM}
\end{figure}

\begin{figure}
    \centering
    \includegraphics[width=0.9\textwidth]{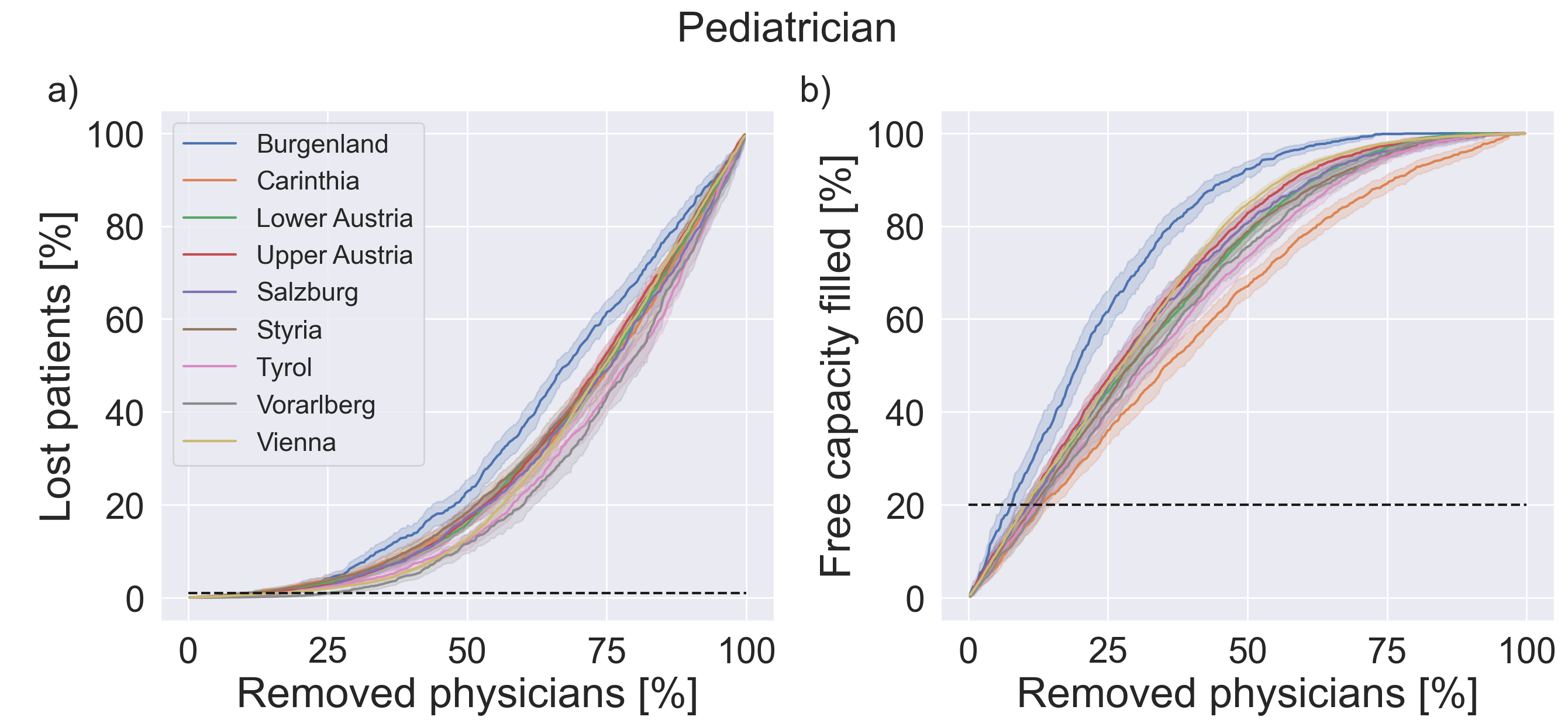}
     \caption{Lost patients and free capacity as a function of removed physicians for pediatricians.}
    \label{fig:PED}
\end{figure}

\begin{figure}
    \centering
    \includegraphics[width=0.9\textwidth]{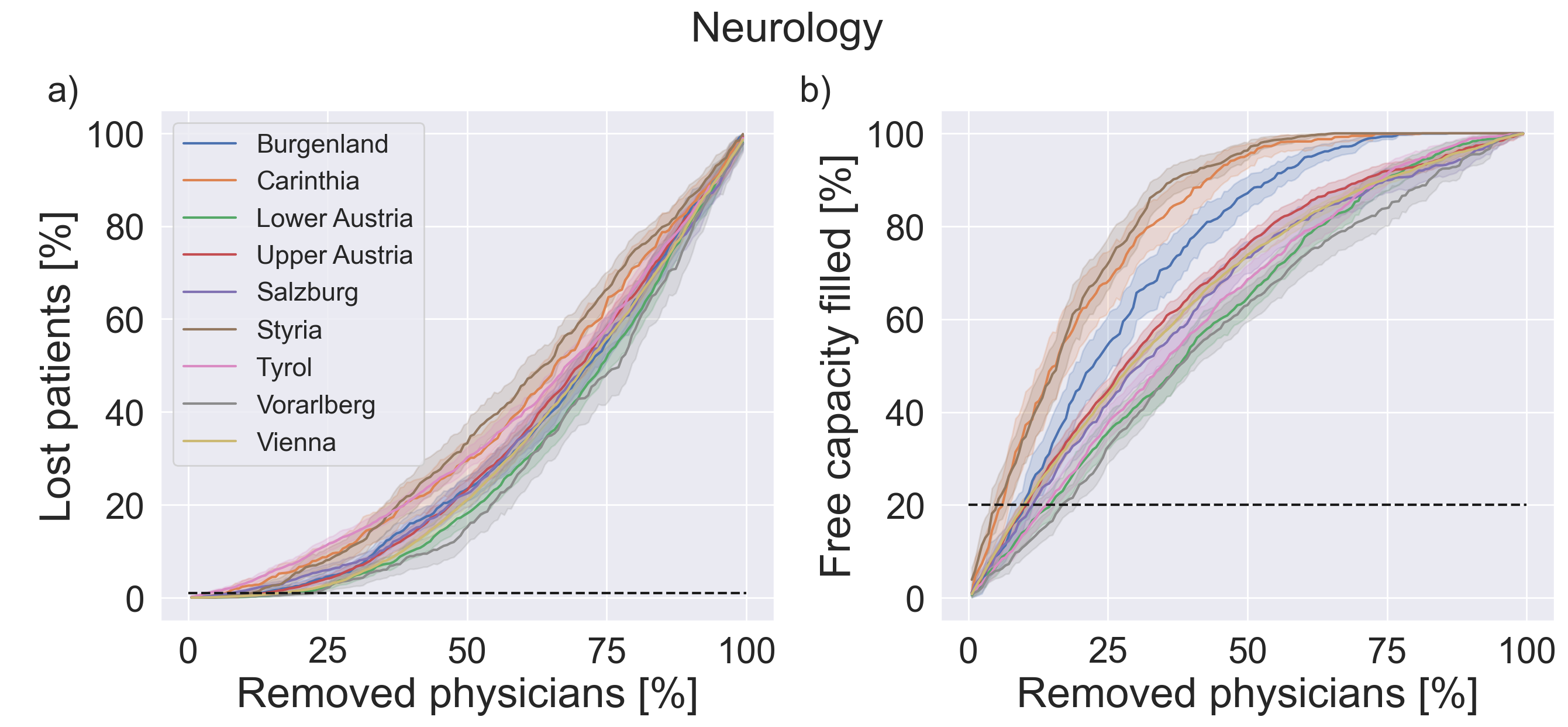}
     \caption{Lost patients and free capacity as a function of removed physicians for neurologists.}
    \label{fig:NEU}
\end{figure}

\begin{figure}
    \centering
    \includegraphics[width=0.9\textwidth]{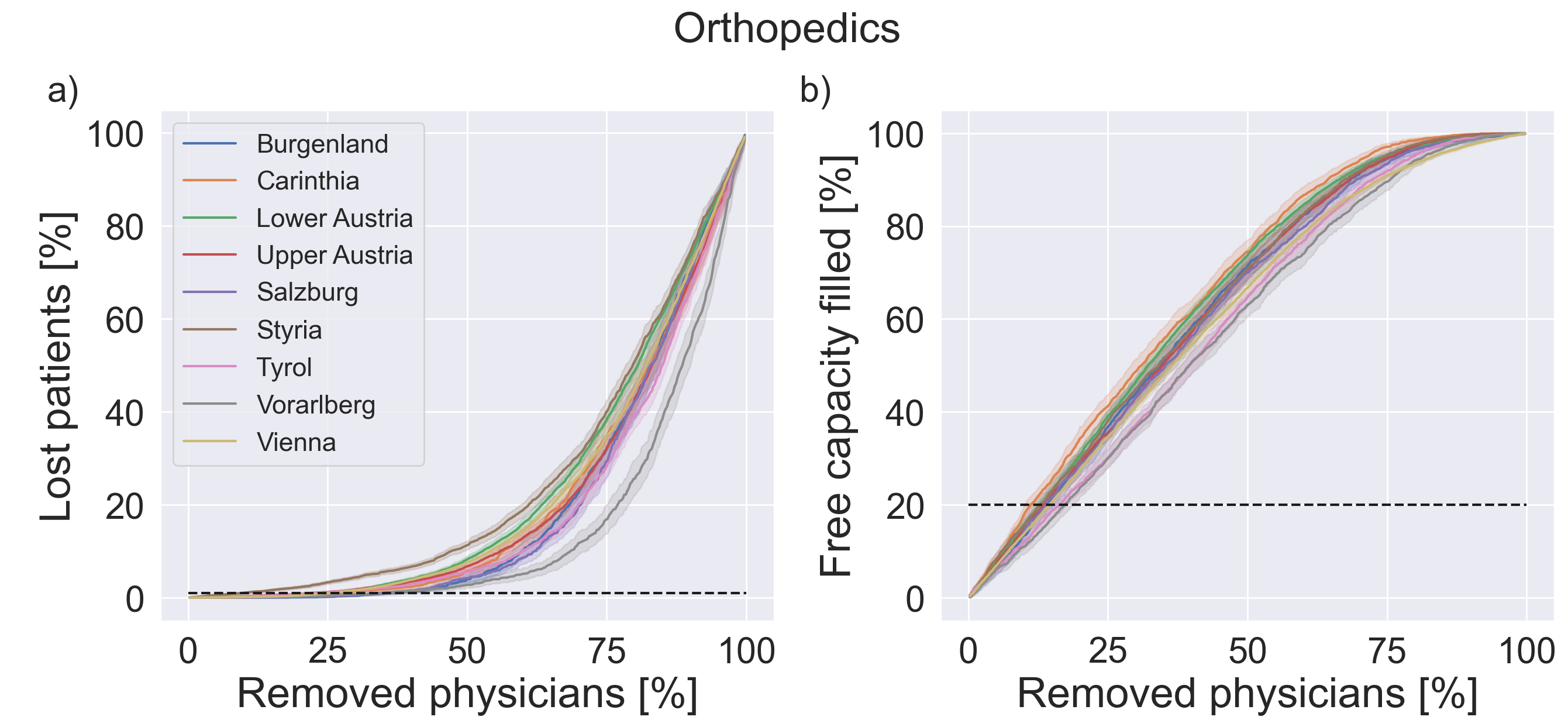}
     \caption{Lost patients and free capacity as a function of removed physicians for orthopedics.}
    \label{fig:ORTH}
\end{figure}

\begin{figure}
    \centering
    \includegraphics[width=0.9\textwidth]{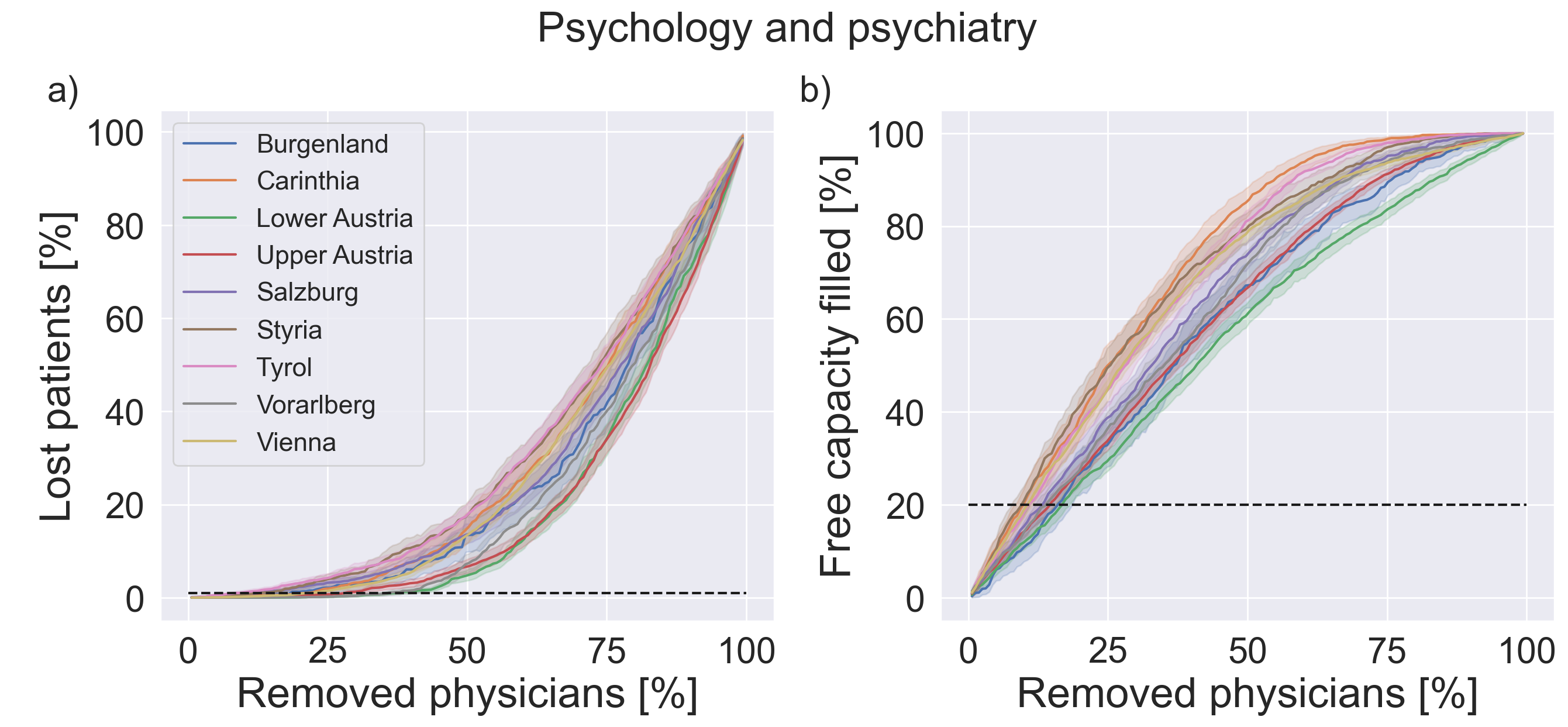}
     \caption{Lost patients and free capacity as a function of removed physicians for psychology and psychiatry.}
    \label{fig:PSY}
\end{figure}

\begin{figure}
    \centering
    \includegraphics[width=0.9\textwidth]{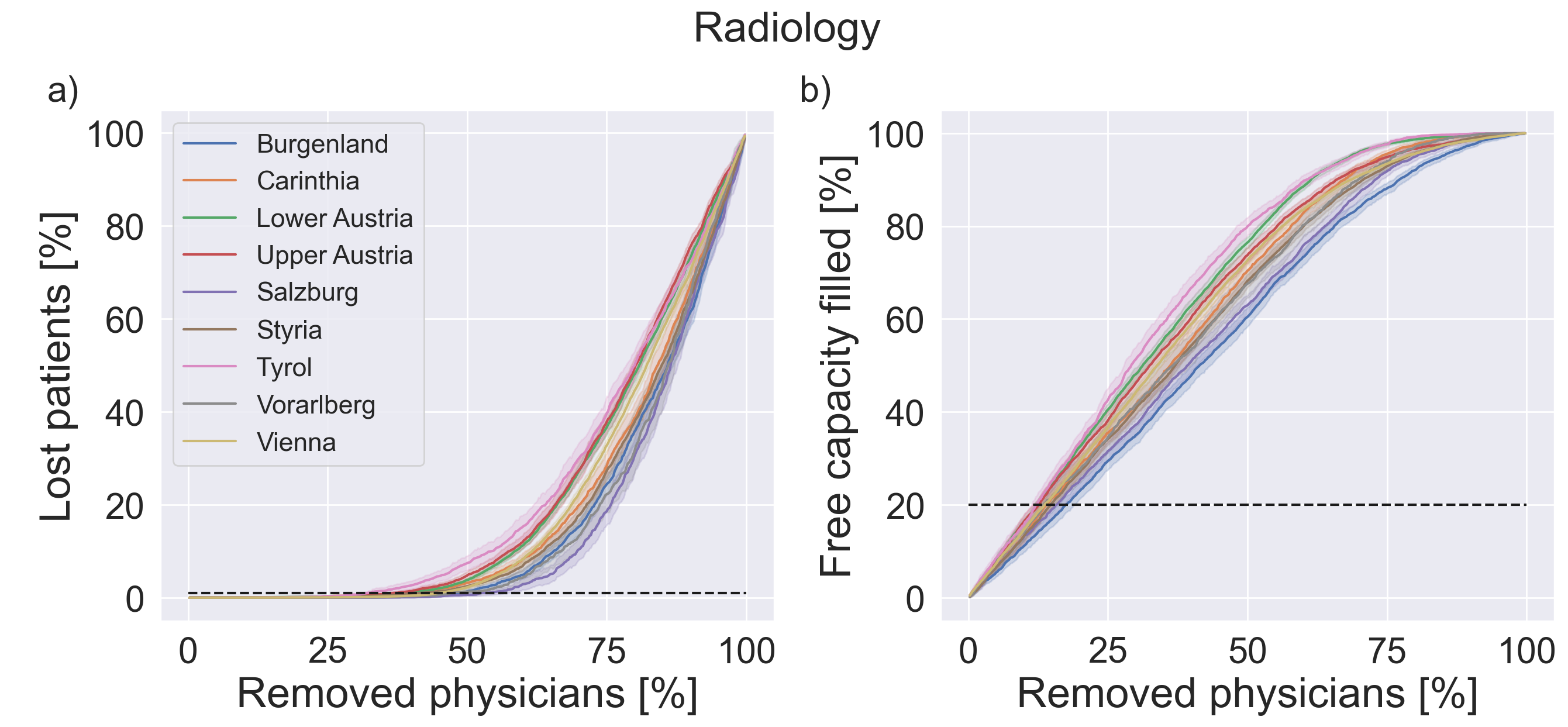}
     \caption{Lost patients and free capacity as a function of removed physicians for radiology.}
    \label{fig:RAD}
\end{figure}

\begin{figure}
    \centering
    \includegraphics[width=0.9\textwidth]{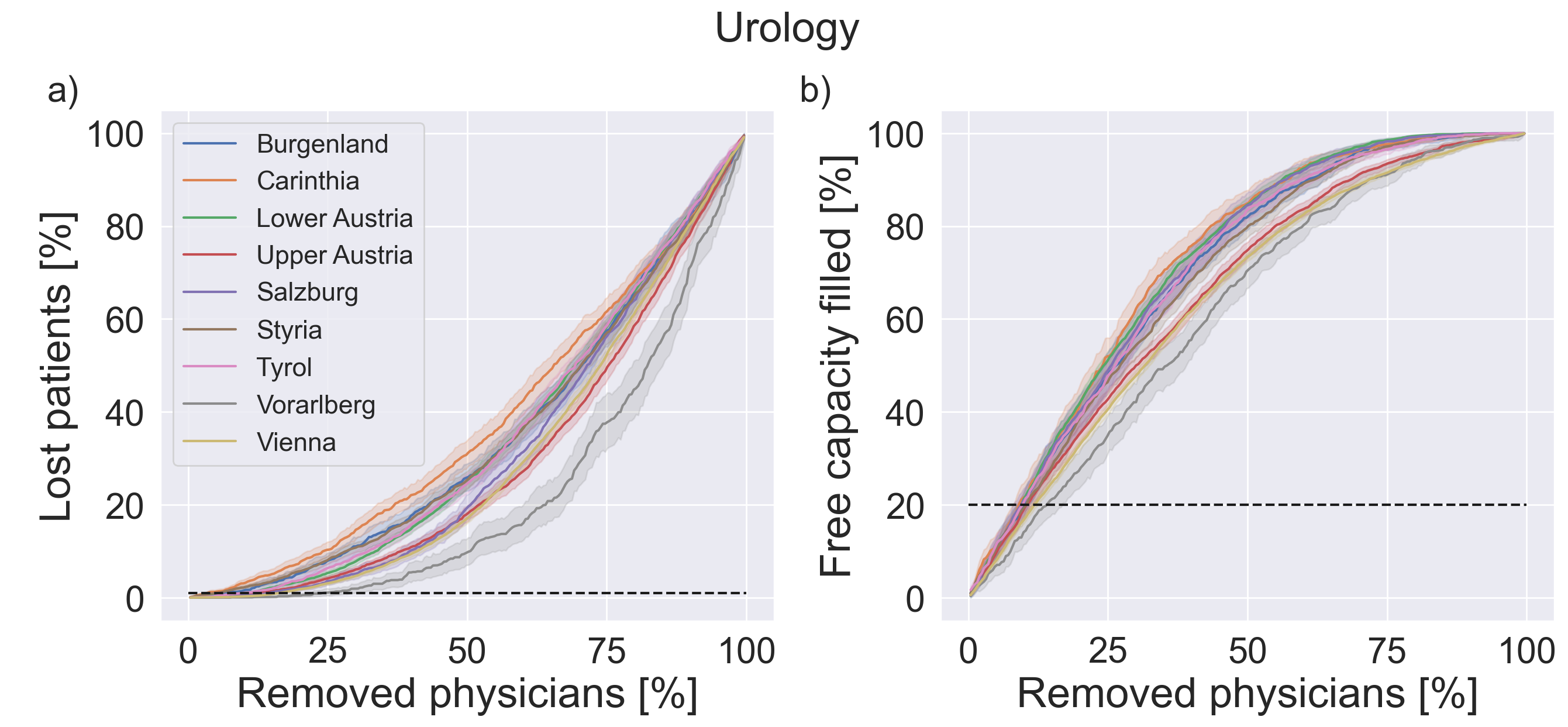}
     \caption{Lost patients and free capacity as a function of removed physicians for urology.}
    \label{fig:URO}
\end{figure}

\newpage
\section{Regional resilience indicators - Heat-map table}
The following tables ~\ref{tab:heatmap_table_LP} and ~\ref{tab:heatmap_table_FC} contain the averaged values and standard deviation over all simulation runs for the resilience indicators $L_{LP}$ and $L_{FC}$ for all federal states and specialists as listed in the heat-map figures in the main text.

\begin{table}
    \centering
    \resizebox{\columnwidth}{!}{\begin{tabular}{l|r|r|r|r|r|r|r|r|r}
     & Burgenland & Carinthia & Lower Austria & Upper Austria & Salzburg & Styria & Tyrol & Vorarlberg & Vienna \\
    \toprule
    GP& 21.0 (2.0) & 28.7 (2.8) & 17.8 (1.4) & 35.3 (2.0) & 30.3 (4.0) & 20.3 (2.0) & 19.4 (3.1) & 43.0 (3.4) & 56.9 (2.0)  \\
    PED& 22.0 (13.5) & 17.9 (12.1) & 14.9 (7.9) & 18.6 (9.7) & 24.1 (11.1) & 13.8 (7.0) & 21.4 (9.9) & 33.7 (12.0) & 17.4 (5.2)  \\
    OBGYN& 17.9 (10.7) & 10.0 (6.5) & 16.0 (5.8) & 13.7 (4.6) & 21.1 (7.6) & 10.1 (4.8) & 19.6 (7.5) & 23.2 (9.8) & 8.4 (3.1)  \\
    OPH& 19.9 (11.3) & 17.8 (9.8) & 7.2 (4.0) & 12.9 (5.4) & 12.6 (8.2) & 7.0 (5.3) & 11.6 (7.4) & 27.6 (14.2) & 11.9 (3.9)  \\
    URO& 17.6 (13.2) & 11.2 (9.4) & 11.7 (6.7) & 15.2 (8.1) & 22.5 (11.7) & 9.4 (7.1) & 16.4 (9.0) & 47.1 (20.3) & 17.6 (4.0)  \\
    DER& 21.1 (13.5) & 16.5 (11.0) & 16.6 (6.8) & 22.4 (7.3) & 23.0 (11.3) & 6.9 (5.6) & 19.1 (10.1) & 37.1 (16.2) & 29.0 (5.0)  \\
    ENT& 22.5 (13.2) & 11.1 (8.2) & 15.6 (6.7) & 21.5 (8.3) & 26.8 (12.6) & 12.9 (8.0) & 26.6 (14.4) & 36.6 (14.5) & 23.0 (6.0)  \\
    IM& 30.7 (8.6) & 26.4 (7.7) & 29.2 (5.9) & 30.4 (7.3) & 36.1 (7.8) & 20.3 (5.5) & 25.0 (7.7) & 52.9 (7.7) & 34.4 (4.5)  \\
    SRG& 46.9 (12.4) & 45.5 (15.3) & 35.7 (9.0) & 42.9 (14.1) & 59.7 (11.5) & 13.8 (10.9) & 29.4 (15.5) & 64.5 (23.9) & 33.2 (7.4)  \\
    ORTH& 43.6 (10.9) & 33.9 (13.7) & 24.7 (8.2) & 25.0 (9.7) & 42.1 (12.4) & 12.9 (8.6) & 27.8 (13.5) & 45.3 (17.5) & 27.2 (5.6)  \\
    NEU& 24.6 (14.3) & 22.0 (18.2) & 25.0 (9.6) & 21.1 (10.3) & 15.8 (12.2) & 25.4 (13.1) & 10.2 (7.1) & 36.8 (20.0) & 18.4 (6.0)  \\
    RAD& 52.5 (6.8) & 45.7 (12.4) & 40.5 (6.8) & 41.5 (9.4) & 61.3 (8.8) & 44.4 (8.7) & 40.2 (11.5) & 59.8 (13.4) & 46.5 (4.7)  \\
    PSY& 44.9 (23.1) & 29.7 (14.4) & 43.1 (10.4) & 38.6 (14.2) & 22.1 (11.5) & 24.0 (14.2) & 19.3 (13.8) & 46.1 (14.4) & 27.8 (9.3)  \\
    \bottomrule
    \end{tabular}}
    \caption{Lost patients indicator. Mean (standard deviation) of resilience indicators $L_{LP}$ of all states and specialties in addition to panel a) of heat-map Fig. 3 in the main text.}
    \label{tab:heatmap_table_LP}
\end{table}

\begin{table}
    \centering
    \resizebox{\columnwidth}{!}{\begin{tabular}{l|r|r|r|r|r|r|r|r|r}
     & Burgenland & Carinthia & Lower Austria & Upper Austria & Salzburg & Styria & Tyrol & Vorarlberg & Vienna \\
    \toprule
    GP& 9.4 (2.1) & 9.2 (1.0) & 8.6 (0.7) & 11.6 (1.0) & 9.8 (1.3) & 8.9 (1.2) & 9.7 (1.0) & 12.5 (1.9) & 15.0 (0.7)  \\
    PED& 10.1 (7.0) & 15.5 (8.4) & 11.1 (4.1) & 10.9 (4.9) & 13.7 (7.9) & 13.1 (5.6) & 13.6 (7.2) & 15.5 (7.8) & 10.8 (3.5)  \\
    OBGYN& 10.5 (6.1) & 10.7 (5.1) & 9.9 (3.0) & 9.8 (3.1) & 10.3 (4.9) & 10.4 (3.0) & 14.2 (5.0) & 14.3 (6.2) & 13.2 (2.7)  \\
    OPH& 10.9 (7.2) & 11.7 (7.1) & 10.9 (4.2) & 11.4 (4.2) & 13.2 (5.5) & 11.1 (4.1) & 13.2 (6.6) & 13.7 (7.1) & 10.7 (2.8)  \\
    URO& 11.4 (6.8) & 11.3 (8.9) & 9.6 (3.3) & 11.5 (4.6) & 10.8 (7.5) & 10.9 (6.0) & 11.0 (7.0) & 22.0 (11.2) & 12.1 (2.8)  \\
    DER& 11.5 (8.9) & 8.2 (5.9) & 12.0 (4.1) & 10.6 (4.2) & 12.4 (6.9) & 11.9 (6.2) & 12.0 (6.8) & 18.1 (10.3) & 11.4 (2.9)  \\
    ENT& 11.9 (7.7) & 10.1 (7.1) & 9.8 (3.7) & 9.9 (4.0) & 14.1 (7.0) & 11.5 (5.0) & 14.0 (5.6) & 13.8 (9.2) & 11.4 (3.4)  \\
    IM& 13.6 (5.3) & 14.4 (4.6) & 13.4 (3.8) & 12.3 (3.7) & 14.6 (4.7) & 13.6 (2.9) & 14.4 (3.7) & 15.0 (4.2) & 15.3 (3.5)  \\
    SRG& 14.8 (9.2) & 16.4 (8.5) & 14.9 (5.1) & 16.7 (5.2) & 17.1 (7.6) & 13.0 (7.8) & 17.0 (8.9) & 22.3 (18.5) & 14.9 (4.5)  \\
    ORTH& 15.1 (7.3) & 12.7 (5.7) & 13.4 (3.6) & 13.9 (3.9) & 15.0 (6.4) & 14.4 (7.2) & 18.1 (7.3) & 17.9 (7.4) & 14.5 (3.2)  \\
    NEU& 15.2 (10.6) & 11.7 (10.0) & 19.2 (11.7) & 11.0 (6.0) & 14.3 (9.8) & 9.8 (8.0) & 15.3 (9.4) & 21.7 (11.3) & 10.3 (3.9)  \\
    RAD& 18.5 (6.8) & 15.4 (6.3) & 12.8 (3.2) & 13.5 (5.7) & 16.9 (7.5) & 14.9 (3.8) & 13.0 (5.9) & 15.6 (8.7) & 14.0 (3.0)  \\
    PSY& 19.7 (13.9) & 14.7 (9.4) & 18.5 (7.8) & 15.5 (5.8) & 15.6 (8.8) & 12.4 (8.1) & 12.4 (7.5) & 16.1 (10.1) & 11.5 (5.0)  \\
    \bottomrule
    \end{tabular}}
    \caption{Free capacity indicator. Mean (standard deviation) of resilience indicators $L_{LP}$ of all states and specialties in addition to panel b) of heat-map Fig. 3 in the main text.}
    \label{tab:heatmap_table_FC}
\end{table}

\newpage
\section{Interactive online visualisation tool}
\begin{figure*}[htb]
    \centering
    \includegraphics[width=1\textwidth]{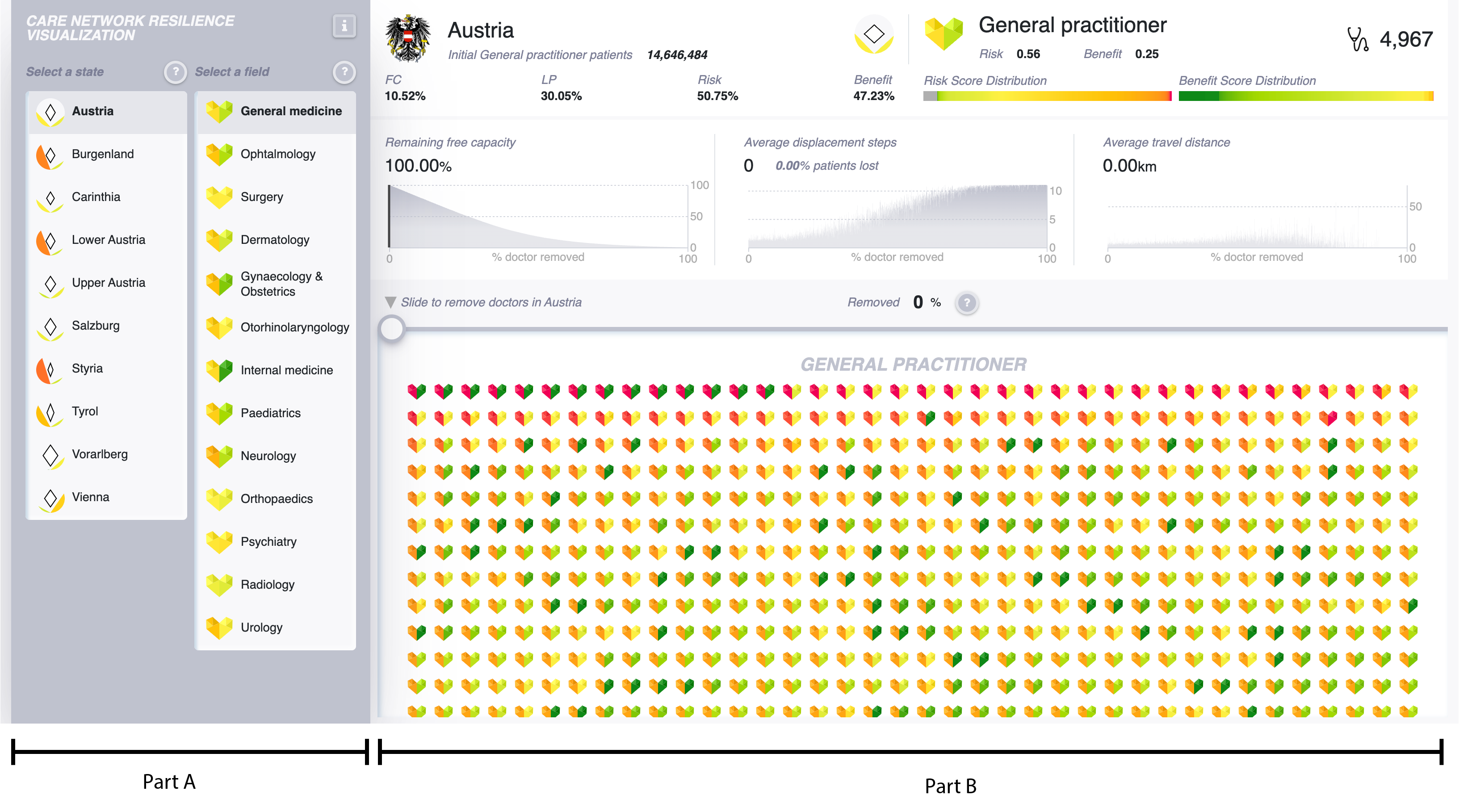}
    \caption{Overview of the online tool's interface. A) overview of resilience indicators; B) result view of the selection in A), currently showing detailed indicators and all physicians for general medicine in Austria .}
    \label{fig:tool_landing}
\end{figure*}

\begin{figure*}[htb]
    \centering
    \includegraphics[width=1\textwidth]{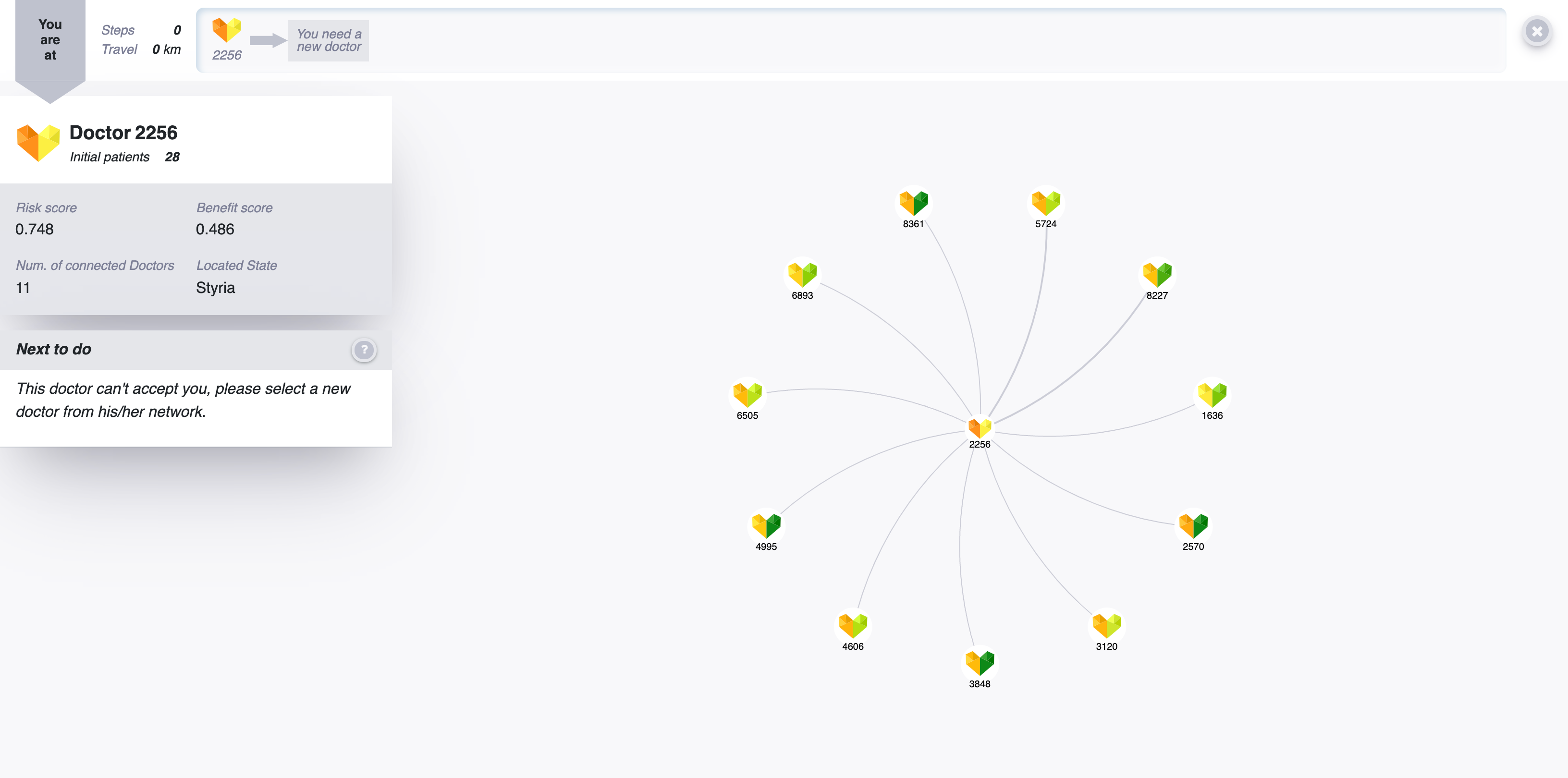}
    \caption{A screenshot of the detailed network of a selected physician. The strength of the connections can be used to compare how many patients two physicians shared relatively to the other physicians in the network. Detailed values can be retrieved by hovering with the mouse pointer over a connection.}
    \label{fig:tool_landing2}
\end{figure*}

The tool contains three components (see Fig.  ~\ref{fig:tool_landing} and ~\ref{fig:tool_landing2}): 1) an overview of resilience indicators per federal state and medical specialty; 2) a detailed view of indicators for the selected specialty in the selected federal state, including risk \& benefit scores, information on free capacity \& lost patients; 3) the physicians’ network within a medical field, displaying indicators per physician. 

In order to allow the comparison of resilience indicator on an aggregate level, we developed two custom glyphs that respectively summarise the characteristics of each federal state and medical specialty. 

\begin{figure*}[htb]
    \centering
    \includegraphics[width=0.7\textwidth]{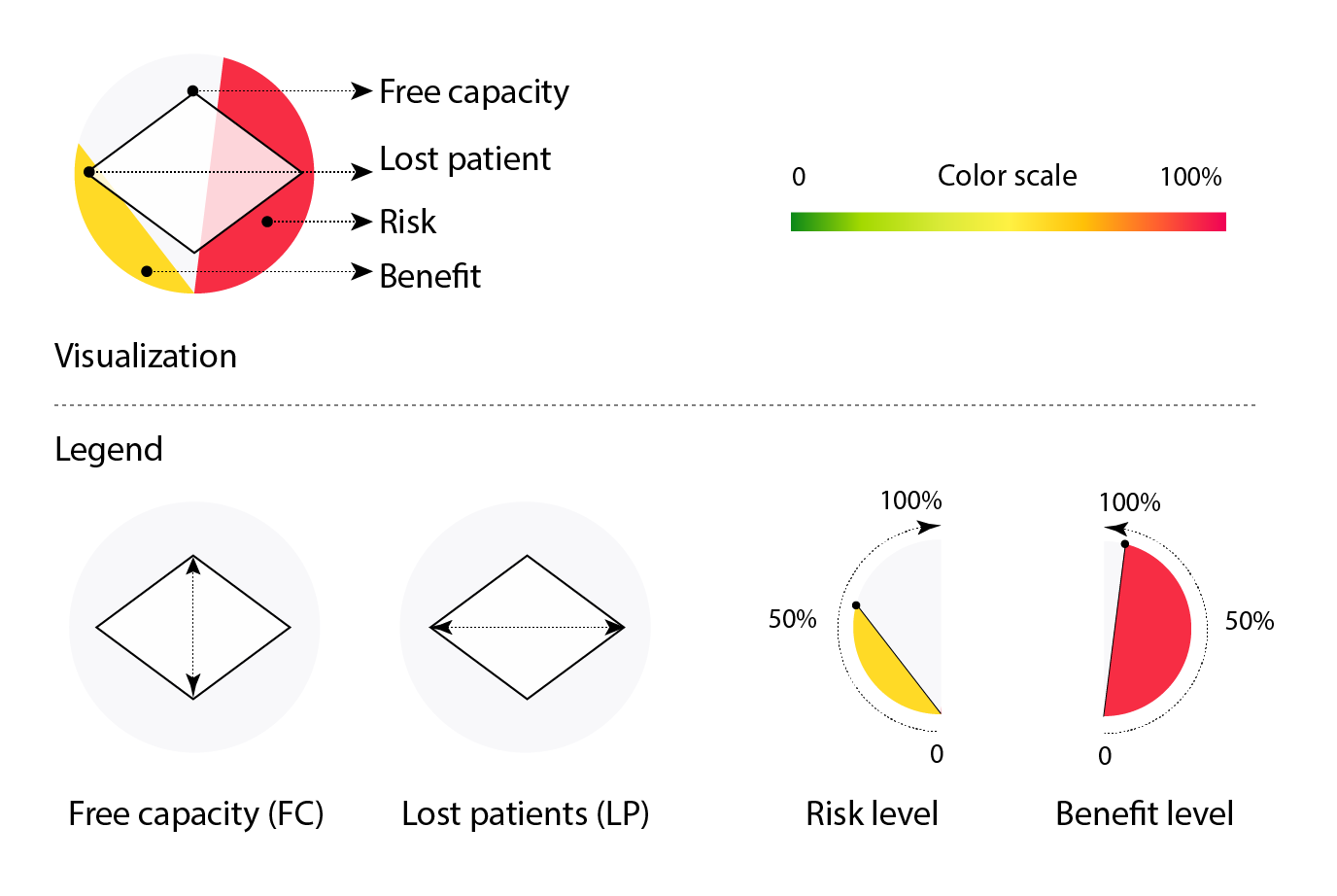}
    \caption{The custom glyph encoding four aggregated indicators per federal state.}
    \label{fig:tool_legend_regional}
\end{figure*}

To visualise a federal state's resilience, we consider four attributes: free capacity, lost patients, risk level and benefit level. The values of free capacity and lost patients are the average values of all the medical specialties in the federal state. In the glyph shown in Fig. ~\ref{fig:tool_legend_regional}, free capacity and lost patients values are respectively represented by the length of the diagonals of a rhombus. Increased length represents a higher value, so that the larger area of the rhombus implies a better resilience performance as physicians are removed from the system. At the same time, the dimensions of the rhombus provide visual cues of the proportion of these two attributes. The values of risk and benefit on the federal state level are the percentage of physicians per state with a risk/benefit score that is higher than the nationwide average risk/benefit score. The risk and benefit attributes are respectively represented by a colour-coded filled arc area in a semicircle.  Green thereby means a low value, whereas red indicates a high value.

\begin{figure*}[htb]
    \centering
    \includegraphics[width=0.7\textwidth]{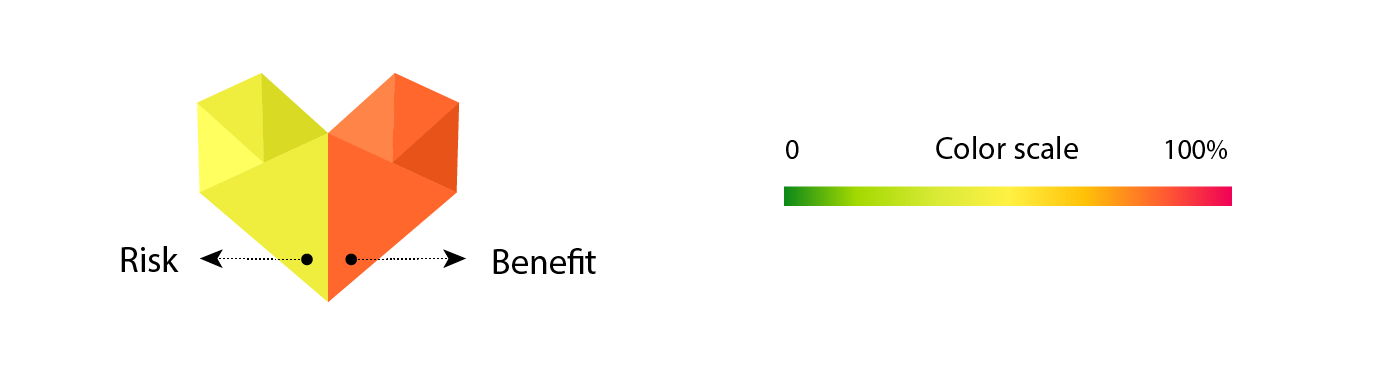}
    \caption{The custom glyph encoding an aggregate of a medical specialty's resilience indicators. The same glyph is used to represent individual physicians.}
    \label{fig:tool_legend_physician}
\end{figure*}

To encode the aggregate resilience of a medical specialty, we present the average values of the risk and benefit scores of physicians in a field within the two halves of a heart shaped glyph depicted in Fig.  ~\ref{fig:tool_legend_physician}. We use a similar approach as in the semicircle representation for federal states: the colour of the left half of heart shows the risk score and the right half shows the benefit score. 

\begin{figure*}[htb]
    \centering
    \includegraphics[width=1\textwidth]{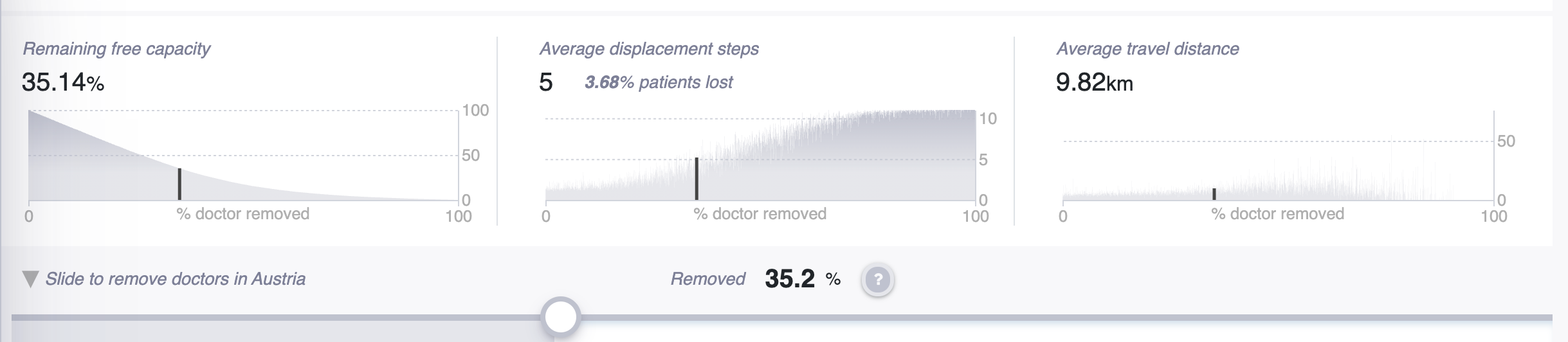}
    \caption{The three charts of the physician removal model results indicating values for a physician removal rate of 35.2\%. The resulting free capacity is 34.15\%, patients take an average of 5 steps and travel 9.82km to find a physician; 3.68\% of patients are lost.}
    \label{fig:tool_removephysician}
\end{figure*}

The overview panel of resilience indicators also serves as a filter for selecting a particular medical field in a particular federal state. After the selection, three sub-components are provided: First, a basic profile that contains four attributes on state level (free capacity, lost patients, risk and benefit levels) and two attributes on specialty level (risk and benefit scores) that are described as an aggregate as well as by their distributions, and finally the number of physicians and patients. The second component shows the model results during a physician removal step (Fig. ~\ref{fig:tool_removephysician}). The third component displays the list of physicians in the selected specialty.

The model results contain three charts: remaining free capacity, average displacement steps and average travel distance. The charts visualise the value changes in respect to the percentage of removed physicians. Users can use a slider to adjust the value from "0\%" to "100\%", in order to remove a certain amount of physicians (on national level) and compare the values.

\begin{figure*}[htb]
    \centering
    \includegraphics[width=1\textwidth]{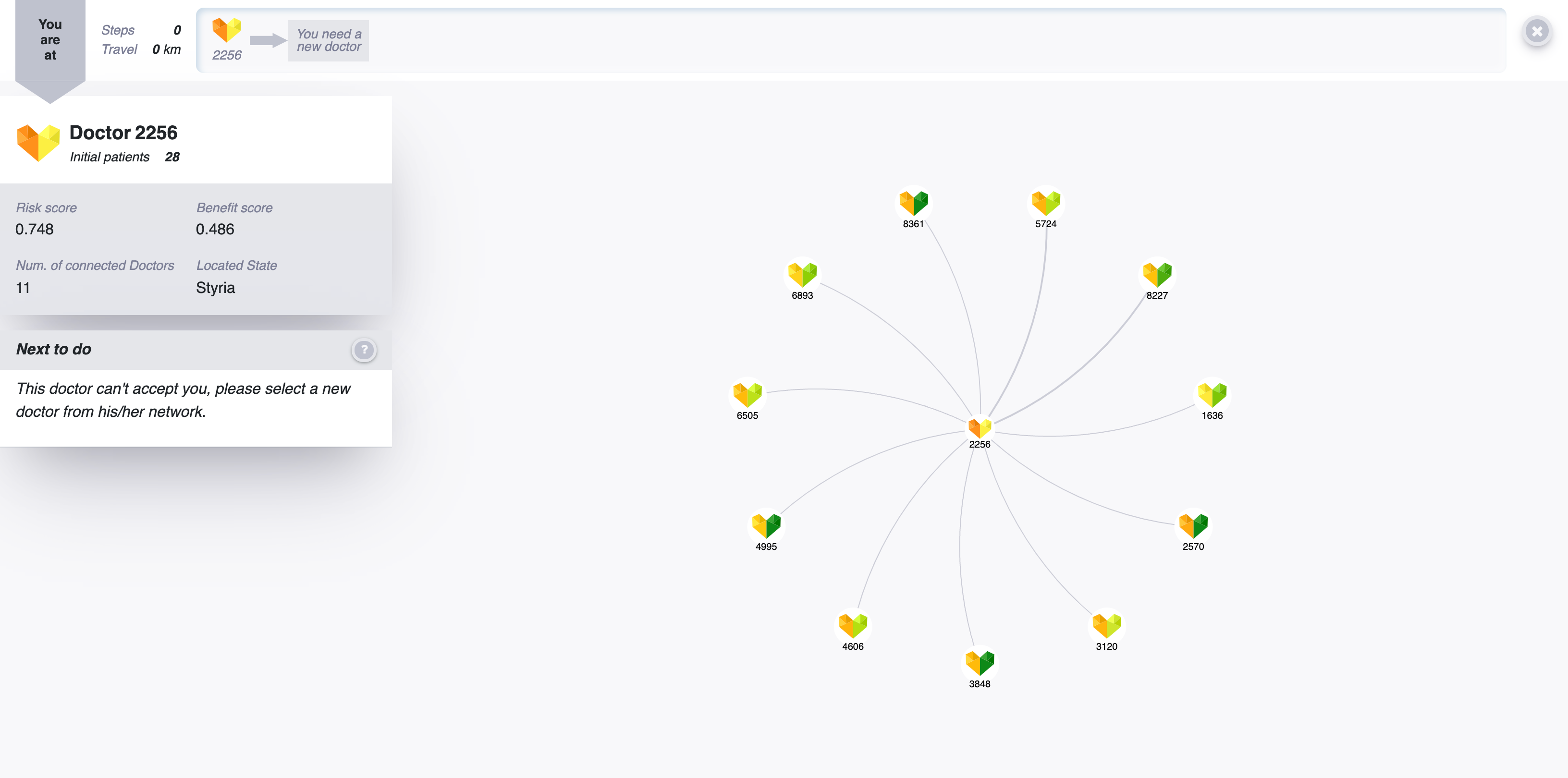}
    \caption{Selecting a physician from the list opens a pop-up window on the left, displaying the profile of the selected physician.}
    \label{fig:tool_physician_view}
\end{figure*}

Below the model results panel, the list of physicians in the selected field are displayed. Users can click on a physician and view the physician's profile as viewed in Fig. ~\ref{fig:tool_physician_view}. The profile displays the physician's state, risk and benefit scores, the number of initial and shared patients, and a snap-view of the physician's network. Clicking the explore button, leads to the third component of the tool: physicians’ networks.

\begin{figure*}[htb]
    \centering
    \includegraphics[width=0.5\textwidth]{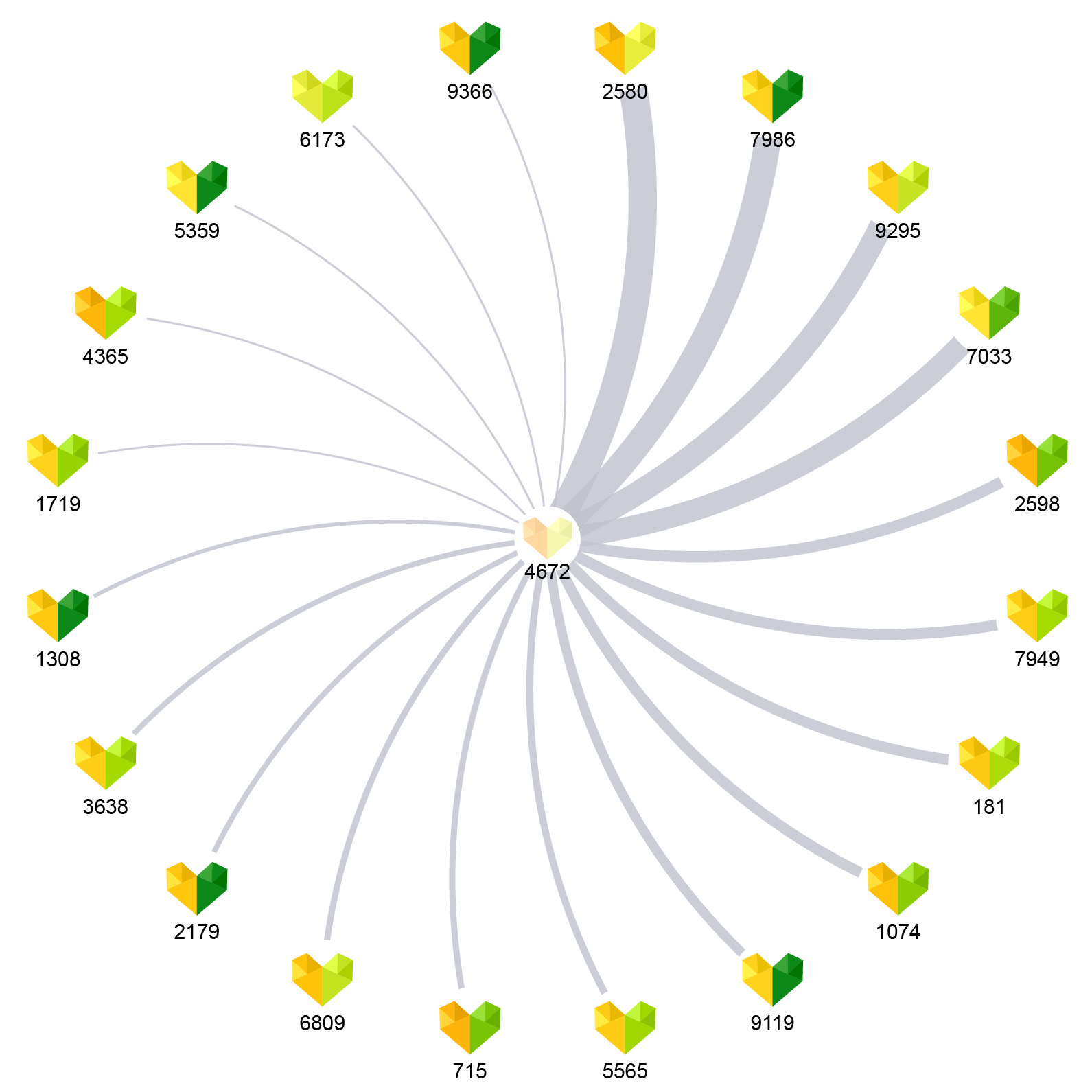}
    \caption{A screenshot of a selected physician's network: the nodes represent the connected physicians, i.e., physicians who have shared patients with the current physician(ID: 4672). The width of edges represents the number of shared patients. Wider edges indicate a higher number.}
    \label{fig:tool_network}
\end{figure*}

The physicians’ network component is used to illustrate the displacement steps of a "lost" patient. It assumes that the current chosen physician always turns out to be unavailable, so the user has to choose another physician based on the current physician's network (see Fig. ~\ref{fig:tool_network}). We expect the user can learn the fundamentals of this simulation system better by interacting with a visual network.

\begin{figure*}[htb]
    \centering
    \includegraphics[width=0.8\textwidth]{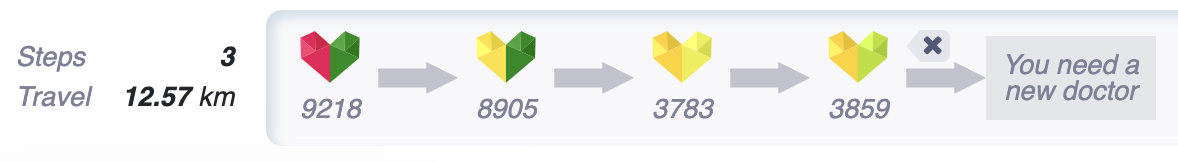}
    \caption{A screenshot of the physician visit history: in this case the user has visited three additional physicians and travelled 12.57 km from the starting position.}
    \label{fig:tool_visit_hisotry}
\end{figure*}

In the visit history in Fig. ~\ref{fig:tool_visit_hisotry} on the top of the panel, users can view the steps they have made and undo the previous step. Since in our model we also assume that patients can (and are willing to) travel only a limited distance from their starting location to find a new physician and is only willing to contact a limited number of new physicians, the user cannot select a physician who is more than 100 km away from the starting location of their initial physician, or over 10 steps away in the physician network. As such, the user always plays the role of a "lost" patient in order to learn how patients can get lost in the system. 

\end{document}